\documentclass[numberedappendix]{emulateapj}
\usepackage{apjfonts, natbib}
\usepackage{rotating}
\usepackage{lscape}

\newcommand{\be}{\begin{equation}}
\newcommand{\ee}{\end{equation}}

\newcommand{\I}{\it{I}\rm}
\newcommand{\Q}{\it{Q}\rm}
\newcommand{\U}{\it{U}\rm}

\newcommand{\flow}{100~GHz}
\newcommand{\fhigh}{150~GHz}
\newcommand{\HII}{{\sc H\,ii}} 

\shorttitle{QUaD Galactic Plane - Compact Sources}
\shortauthors{QUaD collaboration}

\begin{document}

\slugcomment{Submitted to ApJS}

\title{The QUaD Galactic Plane Survey II: A Compact Source Catalog}

\author{
  QUaD collaboration
  --
  T.\,Culverhouse\altaffilmark{1,2},
  P.\,Ade\altaffilmark{3},
  J.\,Bock\altaffilmark{4,5},
  M.\,Bowden\altaffilmark{3,6},
  M.\,L.\,Brown\altaffilmark{7,8},
  G.\,Cahill\altaffilmark{9},
  P.\,G.\,Castro\altaffilmark{10,11},
  S.\,Church\altaffilmark{6},
  R.\,Friedman\altaffilmark{2},
  K.\,Ganga\altaffilmark{11},
  W.\,K.\,Gear\altaffilmark{3},
  S.\,Gupta\altaffilmark{3},
  J.\,Hinderks\altaffilmark{6,12},
  J.\,Kovac\altaffilmark{5},
  A.\,E.\,Lange\altaffilmark{5},
  E.\,Leitch\altaffilmark{4,5},
  S.\,J.\,Melhuish\altaffilmark{3,13},
  Y.\,Memari\altaffilmark{7},
  J.\,A.\,Murphy\altaffilmark{9},
  A.\,Orlando\altaffilmark{3,5}
  C.\,Pryke\altaffilmark{1},
  R.\,Schwarz\altaffilmark{1},
  C.\,O'\,Sullivan\altaffilmark{9},
  L.\,Piccirillo\altaffilmark{3,13},
  N.\,Rajguru\altaffilmark{3,14},
  B.\,Rusholme\altaffilmark{6,15},
  A.\,N.\,Taylor\altaffilmark{7},
  K.\,L.\,Thompson\altaffilmark{6},
  A.\,H.\,Turner\altaffilmark{3},
  E.\,Y.\,S.\,Wu\altaffilmark{6}
  and
  M.\,Zemcov\altaffilmark{3,4,5}
}

\altaffiltext{1}{Kavli Institute for Cosmological Physics,
  Department of Astronomy \& Astrophysics, Enrico Fermi Institute, University of Chicago,
  5640 South Ellis Avenue, Chicago, IL 60637, USA.}
\altaffiltext{2}{{\em Current address}: Owens Valley Radio Observatory, Big Pine, CA 93513}
\altaffiltext{3}{School of Physics and Astronomy, Cardiff University,
  Queen's Buildings, The Parade, Cardiff CF24 3AA, UK.}
\altaffiltext{4}{Jet Propulsion Laboratory, 4800 Oak Grove Dr.,
  Pasadena, CA 91109, USA.}
\altaffiltext{5}{California Institute of Technology, Pasadena, CA
  91125, USA.}
\altaffiltext{6}{Kavli Institute for Particle Astrophysics and
Cosmology and Department of Physics, Stanford University,
382 Via Pueblo Mall, Stanford, CA 94305, USA.}
\altaffiltext{7}{Institute for Astronomy, University of Edinburgh,
  Royal Observatory, Blackford Hill, Edinburgh EH9 3HJ, UK.}
\altaffiltext{8}{{\em Current address}: Cavendish Laboratory,
  University of Cambridge, J.J. Thomson Avenue, Cambridge CB3 OHE, UK.}
\altaffiltext{9}{Department of Experimental Physics,
  National University of Ireland Maynooth, Maynooth, Co. Kildare,
  Ireland.}
\altaffiltext{10}{{\em Current address}: CENTRA, Departamento de F\'{\i}sica,
  Edif\'{\i}cio Ci\^{e}ncia, Piso 4,
  Instituto Superior T\'ecnico - IST, Universidade T\'ecnica de Lisboa,
  Av. Rovisco Pais 1, 1049-001 Lisboa, Portugal.}
\altaffiltext{11}{Laboratoire APC/CNRS, B\^atiment Condorcet,
  10, rue Alice Domon et L\'eonie Duquet, 75205 Paris Cedex 13, France.}
\altaffiltext{12}{{\em Current address}: NASA Goddard Space Flight
  Center, 8800 Greenbelt Road, Greenbelt, Maryland 20771, USA.}
\altaffiltext{13}{{\em Current address}: School of Physics and
  Astronomy, University of
  Manchester, Manchester M13 9PL, UK.}
\altaffiltext{14}{{\em Current address}: Department of Physics and Astronomy, University
  College London, Gower Street, London WC1E 6BT, UK.}
\altaffiltext{15}{{\em Current address}:
  Infrared Processing and Analysis Center,
  California Institute of Technology, Pasadena, CA 91125, USA.}

\begin{abstract}

We present a catalog of compact sources derived from the QUaD Galactic Plane
Survey.
The survey covers $\sim800$ square degrees of the inner galaxy ($|b|<4^{\circ}$) 
in Stokes \I, \Q, and \U\ parameters at 100 and \fhigh, with angular resolution 
5 and 3.5 arcminutes respectively.
505 unique sources are identified in \I, of which 239 are spatially 
matched between frequency bands, with 50 (216) detected at 100 (150)~GHz alone; 
182 sources are identified as ultracompact \HII\ (UC\HII) regions.
Approximating the distribution of total intensity source fluxes as a power-law, we find a slope
of $\gamma_{S,100}=-1.8\pm0.4$ at \flow, and $\gamma_{S,150}=-2.2\pm0.4$ at \fhigh.
Similarly, the power-law index of the source two-point angular correlation function
is $\gamma_{\theta,100}=-1.21\pm0.04$ and $\gamma_{\theta,150}=-1.25\pm0.04$.
The total intensity spectral index distribution peaks at $\alpha_{I}\sim0.25$,
 indicating that dust emission is not the only source of radiation produced by these
 objects between 100 and \fhigh; free-free radiation is likely significant
in the \flow\ band.
Four sources are detected in polarized intensity \it{P}\rm, of which three have
matching counterparts in \I.
Three of the polarized sources lie close to the galactic center, Sagittarius A*,
Sagittarius B2 and the Galactic Radio Arc, while the fourth is RCW 49, a bright \HII\ region.
An extended polarized source, undetected by the source extraction algorithm on
account of its $\sim0.5^{\circ}$ size, is identified visually, and is an isolated
example of large-scale polarized emission oriented distinctly from the bulk galactic
dust polarization.

\end{abstract}


\section{Introduction}
\setcounter{footnote}{0}

\begin{figure*}[ht]
\resizebox{\textwidth}{!}{\includegraphics{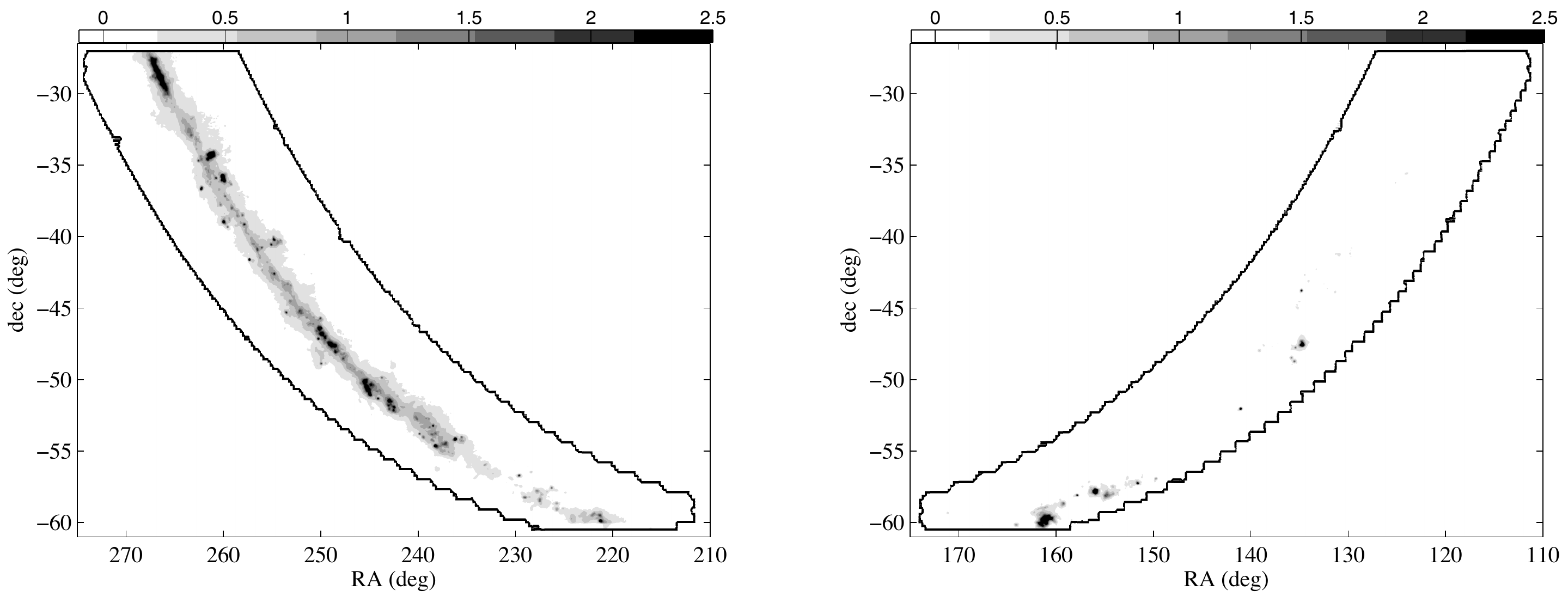}}
\caption{Fourth (left) and third (right) quadrant field-differenced 
\flow\ Stokes \I\ map smoothed to the beam scale ($5'$), with color 
scale in MJy/sr.
The solid black lines indicate the survey coverage.
}
\label{fig:fdmap}
\end{figure*}

Millimeter (mm), sub-millimeter (sub-mm) and far-infrared (FIR) observations are ideal for studying
the properties of star-forming regions in the galaxy, in particular the cool
envelopes of dust and gas which host sites of potential and active star formation.
By spanning the peak in the spectra of these objects, measurements between the mm
and FIR can tightly constrain the parameters of the thermal radiation produced by the dust.
In particular, the mass of a star-forming core and its surrounding envelope is well-traced
by its measured flux in these bands, since this radiation is optically thin at sub-mm and longer
wavelengths.

Surveys covering large sections of the galaxy have the potential to collect statistical
samples of cores in a range of evolutionary states, comparatively free of bias introduced by
targetting particular regions.
These surveys are ideal to study processes related to star-forming regions, such as measuring the
core mass function (from which the stellar initial mass function may be derived), particularly
at the high-mass end, which, on account of the short-lived high mass cores, is understudied 
relative to lower masses~\citep[e.g.][]{enoch2006,young2006,enoch2008}.
Combination with infrared (IR) data yields insight into the ages of cores, permitting 
differentiation between prestellar sub-mm cores, which lack an IR counterpart, and protostellar 
cores, in which the ultraviolet radiation produced by protostars is re-radiated into the mm, sub-mm and
IR by the surrounding envelope.
Phenomena associated with later evolutionary phases, such as mass ejection, dissipation of the envelope, 
and dynamical interactions are not significant in the prestellar or protostellar stage 
--- the mass and spatial distribution of such cores therefore capture information regarding 
the fragmentation process~\citep{enoch2006}.

Observations of polarized radiation permit a window to study the role of magnetic 
fields~\citep[e.g.][]{greaves1995,novak1997}, and their role in providing support against 
collapse.
In the mm and sub-mm, polarization is due to emission along the long axis of dust grains 
partially aligned by the magnetic field, and thus measurements of the dust polarization
directly probe local magnetic fields~\cite[e.g.][]{hildebrand1988}. 
These fields are thought to strongly influence the evolution of molecular clouds, since 
they provide support preventing the collapse of the gas and subsequent triggering of star 
formation.

Several large-scale surveys are already underway or completed to help address these
questions.
Herschel~\citep{Herschel} and Planck~\citep{Planck} will provide extensive spectral coverage from the radio to the far 
infrared, fully characterizing the spectral energy distribution (SED) of star-forming 
cores over the fully sky; selected existing results in targetted regions 
include~\citep[e.g.][]{hennemann2010,juvela2010,andre2010}, but are limited to 
total intensity observations.
Ground and balloon instruments also contribute substantially to the literature:
~\cite{schuller2009} present an APEX LABOCA 95 deg$^2$ survey in total intensity 
with resolution of $19.2''$ at 353 GHz, with the final survey coverage expected to reach
350 deg$^{2}$; Bolocam has mapped 150 deg$^{2}$ of the first galactic quadrant at 1.1mm 
(268~GHz) with resolution $33''$, with a source catalog presented in~\cite{rosolowsky2009}; 
BLAST~\citep{olmi2009,netterfield2009} provide a 50 deg$^{2}$ survey of the Vela molecular 
cloud at 250, 350 and 500 microns (36, 42 and 60 arcsec resolution respectively).

Observations at comparable resolution are currently scarce at $\sim$\flow, and yet 
provide additional constraining power to the Rayleigh-Jeans
tail of the thermal dust spectrum, and probe for contributions due to other emission 
mechanisms which contribute increasingly at lower frequencies (e.g. free-free).
Furthermore, there is little high angular resolution polarization data at these
frequencies, despite their utility in understanding star-forming regions.

In this paper we present a catalog of compact sources found in the QUaD
galactic plane survey~\citep{culverhouse2010}, which covers over 
$\sim800$ square degrees of the low-latitude galactic plane at 100 and \fhigh~with 
beam FWHM of $5'$ and $3.5'$ respectively, in Stokes \I, \Q~and \U~parameters
\footnote[1]{The QUaD maps and source catalogs analyzed in this paper are available 
for public download at \tt{http://find.spa.umn.edu:/quad/quad\_galactic/}}.
A survey of this size, frequency and angular resolution can be used to 
investigate the polarized and unpolarized properties of both diffuse emission
and discrete sources.
The QUaD survey was conducted \it{blind}\rm, in that no region was specifically targetted. 
In principle, this allows the construction of statistical samples of cores, 
representative of the distribution of core masses and ages in the galaxy as a whole.
However, we note that at the $\sim$few arcminute resolution of the survey, the maps
 do not generally resolve individual cores:
Dense cores typically have size $\sim 0.1\mathrm{pc}$~\citep[e.g.][]{Williams2000}, hence 
for nominal distances of a few hundred pc, the sources presented here should be considered 
as `clumps' hosting cores rather than individual cores themselves.
In addition to the lack of resolution and accurate clump distances, the contribution 
of free-free emission at \flow\ biases measurements of the flux from the dust 
component; these caveats prevent reliable mass calculation.
Our goals here are therefore to analyze the observed quantities of the sources
 in the survey, rather than infer their physical properties.

Basic information on the instrument, observations and maps is presented in 
Section~\ref{sec:instobs}. 
In Section~\ref{sec:srcextraction}, we describe our algorithm for extracting sources
 in the presence of a diffuse background.
The global properties of the catalog are discussed in Section~\ref{sec:data}, with the 
full catalogs presented in Table~\ref{tab:srccat} (total intensity) and Table~\ref{tab:polsrccat} 
(polarized intensity). 
Discussion and Conclusions are found in Section~\ref{sec:conclusions}.
Extensive simulations, presented in Appendix~\ref{app:sims}, are used to quantify 
the effects of mapmaking and source extraction algorithm on the recovered source
properties.

\begin{figure*}[ht]
\resizebox{\textwidth}{!}{\includegraphics{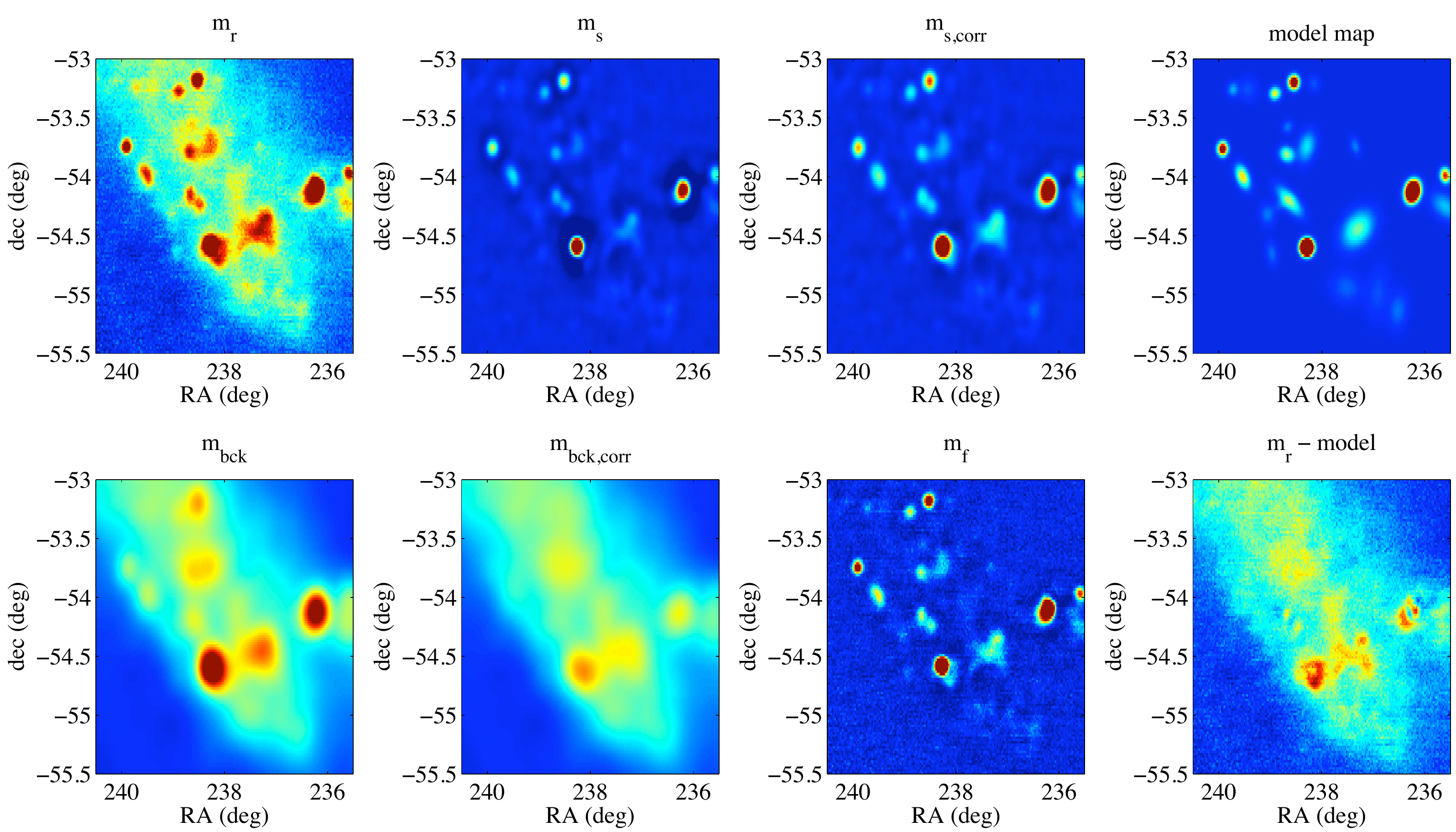}}
\caption{Source segmentation and fitting using a section of 
QUaD \flow\ \I\ data; colorscale is the same 
in all panels and runs from -0.0035 to 0.035 MJy/sr. 
{\it Top Left:} Raw map $m_{r}$.
{\it Bottom Left:} Initial estimate of background map $m_{bck}$.
{\it Top row, second column:} Initial background subtracted map $m_{s}$.
{\it Bottom row, second column:} Background map corrected for discrete 
source flux ($m_{bck,corr}$).
{\it Top row, third column:} Background subtracted map $m_{s}$ using
source-corrected background map $m_{bck,corr}$.
{\it Bottom row, third column:} Map used for source fitting $m_{r}-m_{bck,corr}$.
{\it Top right:} Model of discrete source population using fits to 
$m_{f}$.
{\it Bottom right:} Residual between model sky and input image $m_{r}$. 
}
\label{fig:sourceseg}
\end{figure*}

\section{Instrument, Observations and Maps}
\label{sec:instobs}

Here we summarize the features of the QUaD galactic plane survey.
A detailed description of the instrument can be found in~\cite{hinderks08}, 
hereafter referred to as the ``Instrument Paper''. 
The field selection, survey strategy, data processing and construction of the 
Stokes \I, \Q\ and \U\ maps are presented in~\cite{culverhouse2010},
 hereafter the ``Map Paper''.

QUaD was a 2.6~m Cassegrain radio telescope on the mount originally
constructed for the DASI experiment~\citep{leitch02}. 
This is an az/el mount, with a third axis allowing the entire optics and 
receiver to be rotated around the line of sight.
The mount is enclosed in a reflective ground shield, extended from DASI,
on top of a tower at the MAPO observatory approximately 1~km from
 the geographic South Pole.

The QUaD receiver consisted of 31~pairs of polarization sensitive
bolometers \citep[PSBs;][]{2003SPIE.4855..227J}, 12 at \flow,
and 19 at \fhigh.
The bolometers were read out using AC bias electronics,
and digitized by a 100~Hz, 16~bit ADC; the raw data were staged on 
disk at Pole and transferred out daily via satellite.

The observations reported in this paper were made between July and
October 2007, with the telescope decommissioned in late 2007. 
In total QUaD surveyed the galaxy for 40 days, covering a total of 
$\sim800~\mathrm{deg}^{2}$.
The survey is divided into two regions, approximately covering 
245-295$^\circ$ and 315-5$^\circ$ in galactic longitude $l$, and -4 
to +4$^\circ$ in galactic latitude $b$.
These regions are loosely termed the `third quadrant' and `fourth 
quadrant' throughout.

Maps are made by coadding the timestream from each detector into
flat-sky (RA, dec) pixels of size 0.02$^\circ\times0.02^\circ$.
The absolute pointing accuracy was determined to be $\sim0.5'$ rms,
using pointing checks on RCW 38 and other galactic sources over two
 seasons of CMB observations (see Instrument Paper for further details).
A field-differencing scheme was used to remove spurious ground
contamination; all the results presented here are derived using 
field-differenced maps.
Absolute calibration is applied using a scaling factor at each frequency,
 derived in the QUaD CMB analysis presented in~\cite{QUAD09}.
These factors were calculated by cross-calibrating QUaD CMB maps to those
from the Boomerang experiment~\citep{masi06}, and have an estimated uncertainty
of $3.5\%$.
The \flow\ \I\ map for both survey regions is shown in Figure~\ref{fig:fdmap};
the reader is referred to the Map Paper for similar maps at both
frequencies and in \I, \Q, and \U.
In addition to the sky maps, variance maps for each Stokes parameter 
are also constructed, which give a measure of the noise in each 
map pixel.
The typical survey sensitivity in each survey area is 74 (107) kJy/sr 
at 100 (150) GHz in I, and 98 (120) kJy/sr in \Q/\U, at a 
spatial resolution of 5 (3.5) arcminutes at 100 (150) GHz.
The orientation of $Q$ and $U$ in the QUaD polarization maps follows
the IAU convention~\citep{IAU96}, in which $+Q$ is parallel to N-S and $+U$
 parallel to NE-SW.

\section{Source Extraction}
\label{sec:srcextraction}

As is readily apparent from Figure~\ref{fig:fdmap}, a substantial 
contributor to the sky signal is diffuse emission.
This `background' increases the uncertainty in measured properties of compact 
sources above that due to detector and atmospheric noise.
However, the systematic effect of diffuse emission can be reduced using spatial filtering.

The source extraction method implemented here is an adaptation of the algorithm 
described in~\cite{desert08} (a mexican-hat linear filter in image space), which 
was designed to separate compact sources from the diffuse galactic emission for 
the Archeops experiment.
Maps of each stage of the source extraction algorithm, described below, are 
shown in Figure~\ref{fig:sourceseg} in a representative section of the fourth 
quadrant \flow\ \I\ data.

In our algorithm, two sets of smoothed \I\ and $P=\sqrt{Q^{2}+U^{2}}$ maps are 
made at each frequency; both are derived from the raw maps $m_{r}$.
The first set, $m_{b}$, consists of $m_{r}$ smoothed to the beam scale 
$\sigma_{beam}$.
In the second, $m_{r}$ is smoothed to an angular scale $\sigma_{bck}=2.5\times\sigma_{beam}$ 
to form a template map of the diffuse background, $m_{bck}$.
In both cases, a circularly symmetric gaussian function is the smoothing kernel.
The choice of $\sigma_{bck}$ is designed to minimize the background contribution
to source fluxes without introducing large biases in the measured flux.
In~Appendix~\ref{subsec:sigmabck}, simulations demonstrate the consequences
of other choices of $\sigma_{bck}$ on recovered source fluxes.
The background maps are then subtracted from beam-smoothed maps to yield the
`source extraction' map $m_{s}$:
\begin{equation}
m_{I,s}=m_{I,b}-m_{I,bck}
\label{eq:bcksub}
\end{equation}
\noindent and likewise for $P$.
Note that since $\sigma_{bck}>\sigma_{beam}$, Equation~\ref{eq:bcksub} is
equivalent to convolving $m_{r}$ with filter constructed from
the difference of the two smoothing kernels, commonly referred to as a `mexican hat'
filter.

Negative pixels due to ringing are masked, and the remaining pixels subjected 
to signal-to-noise thresholding.
Pixels above a signal-to-noise threshold of 5 (3) for 
total (polarized) intensity are flagged as belonging to source candidates. 
The polarization data has a lower extraction threshold because the noise
properties are closer to white on account of the unpolarized atmosphere,
and also because the diffuse component amplitude (fractional polarization
$< 2\%$; see Map Paper) is close to the instrumental noise level and therefore
its effect on source fluxes is small.
High signal-to-noise regions in the $P$ map define a set of pixels to which
we fit polarized sources in the \Q\ and \U\ maps separately.
Candidate pixels in all Stokes maps are subject to suitability checks; isolated pixels or
groups of pixels smaller than the beam width are removed.

Sources in close proximity tend to be members of the same thresholded 
region, so an internally-developed segmentation algorithm based
 on the SExtractor code~\citep{bertin96} is used to split such regions into
 separate sources. 
Source segmentation is applied to the \I, \Q, and \U\ maps separately
at each frequency, resulting in a set of six source position lists, along with
the map pixels assigned to each source. 

Having determined source positions, the background maps $m_{bck}$ are 
regenerated by again smoothing $m_{r}$, but with the pixels corresponding 
to discrete sources replaced by the local median --- this `source-corrected' 
background map is denoted $m_{bck,corr}$, with $m_{s,corr}=m_{b}-m_{bck,corr}$
following from Equation~\ref{eq:bcksub}.
The median-replacement step reduces the amount of ringing due to 
the background filtering implemented in Equation~\ref{eq:bcksub} (see Figure~\ref{fig:sourceseg}).
The resulting background map contains less leaked flux due to smoothing 
discrete souces with a kernel larger than the beam size.
The background subtraction, source detection and segmentation stages are
then repeated. 

All ingredients for measuring source properties are present at this point:
the background map $m_{bck,corr}$, a list of pixels 
belonging to each source, and the input map itself $m_{r}$, with its variance 
map $\sigma^{2}_{r}$.
The background map $m_{bck,corr}$ is subtracted from the input map $m_{r}$ yielding
the map to which source models are fit,
\begin{equation}
m_{I,f}=m_{I,r}-m_{I,bck,corr}.
\label{eq:fitmap}
\end{equation}
\noindent and likewise for $P$.
An elliptical gaussian is fit to the resulting pixels for each source; 
an example of the model reconstructed from these fits is shown in Figure~\ref{fig:sourceseg}.
Residuals of this model against $m_{r}$, also shown in Figure~\ref{fig:sourceseg}, indicate that
the source-fitting works well, except for two cases: 1) particularly bright sources, which
can leave residuals at the $\sim\mathrm{few}$ percent level; and 2) sources in close
proximity, for which the source segmentation fails and the sources are classed as a single object.

Fits are performed independently at each frequency in \I, \Q\ and \U, with 
pixel noise taken from the corresponding variance map $\sigma^{2}_{r}$; errors 
on source properties follow directly from the parameter errors returned by the 
fit minimizer. 
The elliptical gaussian fit parameters are used to calculate source properties such 
as flux and position; in a small number of cases, the uncertainty on a fit parameter 
diverges, in which case we do not quote the uncertainty on any physical quantity
 derived from this parameter.
Derived quantities such as spectral index $\alpha=\mathrm{log}(I_{2}/I_{1})/\mathrm{log}(\nu_{2}/\nu_{1})$, 
polarization angle $\phi=0.5\mathrm{tan}^{-1}\left(U/Q\right)$, polarization 
fraction $P/I$, and their associated errors are calculated from \I, \Q\ and \U\ fluxes.
In general, a source detected in \I\ will not have the same set of
pixels as in \Q\ or \U\ as each map is treated independently;
sources are spatially matched across catalogs later to determine e.g. polarization fraction.
Given two catalogs A and B (such as total intensity at two frequencies), each source 
in A is matched to a source in B if their angular separation is less than 
three map pixels ($3.6'$), conservatively larger than the rms day-to-day telescope pointing 
wander of $\sim0.5'$.
If more than one source in B matches a source in A, as can happen when matching sources
between 100 and 150 GHz due to the higher resolution in the latter band, the closer of the two
is selected.
Having matched sources, the corresponding physical quantities are combined to yield 
the derived quantity such as the spectral index.

\subsection{Consequences of Field-differencing}
\label{subsec:fdcons}

\begin{figure}[h]
\resizebox{\columnwidth}{!}{\includegraphics{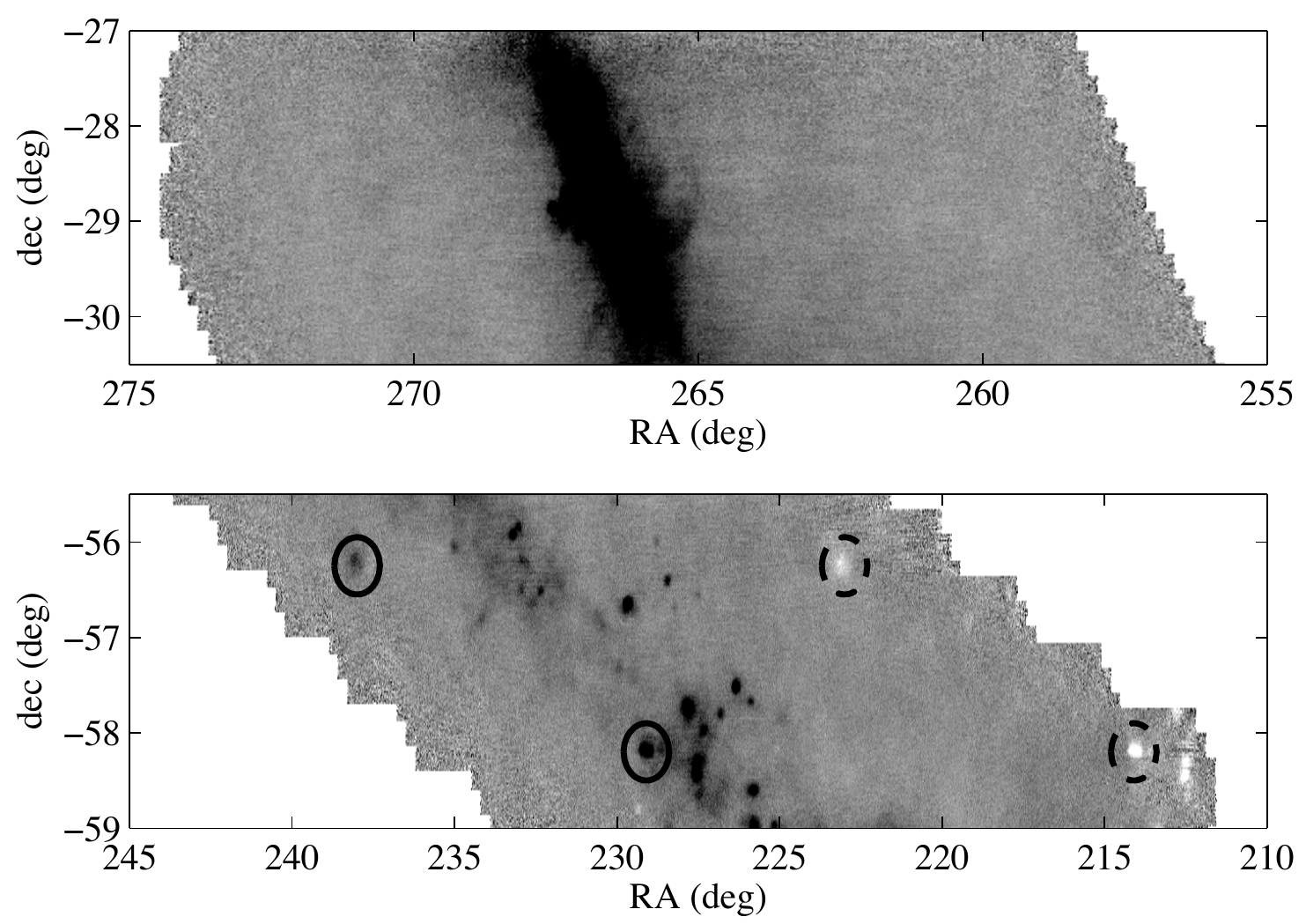}}
\caption{Top: Subsection of field-differenced \flow\ \I\ map covering galactic center.
Few sources lie near the left (high R.A.) edge of the map, where the trail field
is aligned.
Bottom: Subsection of field-differenced \flow\ \I\ map covering the lowest
Decl. of the survey.
Two real sources close to the trail field region (high R.A. edge) are circled with
solid black lines.
Their field-differenced counterparts are circled with broken black lines; notice the
change in sign in intensity of these spurious sources (other field-differenced sources
are also visibile).
No sources lie as close to the trail field edge in the top plot, and hence no
spurious sources are detected at this Decl.
}
\label{fig:fdsrcs}
\end{figure}

The field-differencing operation performed on the timestream to remove ground 
contamination can result in spurious sources in the data.
If a source lies in the trail field of the observations (at larger R.A.), when the 
trail field is differenced against the lead field the source will appear negative
 in the lead field, resulting in a negative measured flux.
Such sources are removed in \I\ by rejecting candidates with fluxes below zero at
a signal-to-noise of 5 or greater.
This rejection is not possible in \Q\ or \U\ because the polarized flux can take
positive or negative values.
Instead, detected polarized sources are matched to the total intensity source catalog;
if the polarized source is matched to a source of negative total intensity with $|\mathrm{S/N}|>5$,
the polarized source is removed from the catalog.
This method may allow small numbers of field differenced polarized sources to leak into
the catalog, since sources in \I\ are extracted at a higher significance threshold ($5\sigma$)
than polarization --- a field differenced source in $P$ may not be matched to a total
intensity candidate and therefore cannot be rejected.

Field-differenced sources are also increasingly expected at low declinations.
The central PSB pair are aligned on the plane of the galaxy, with the low R.A. edge 
of the trail field aligned with the high R.A. edge of the lead field; the width 
of the QUaD focal plane allows for overlapping coverage of the lead and trail fields.
At higher elevation (lower declination) the scan throw of $\delta Az=15^{\circ}$ 
translates into a smaller R.A. range as $\delta\mathrm{R.A.}=\delta\mathrm{Az}\times\cos(\mathrm{Decl.})$.
Due to the alignment of lead and trail field edges and the decreasing scan throw 
in R.A. at lower Decl., the trail field lies closer to 
the galactic plane at low Decl. --- see Figure~\ref{fig:fdsrcs} (we further refer the
 reader to Figure 2 of the Map Paper for a graphical representation of the lead/trail 
field geometry over the full survey).
At lower Decl. the trail field is therefore more likely to contain a bright source 
(Figure~\ref{fig:fdsrcs}).
However, the QUaD catalog indicates that most detected sources lie within three degrees
of the galactic plane (Section~\ref{subsec:bdist}), and thus the contribution of field 
differenced sources is small over most of the survey, since the trail fields never 
get closer than $\sim(15^{\circ}/2)\times\cos(60)=3.75^{\circ}$ to $b=0$.

\section{Results}
\label{sec:data}

Source catalogs from both the third and fourth quadrant maps are extracted 
and combined for the purposes of calculating statistical properties.
Spurious sources are rejected if the flux is negative at one frequency and
 undetected at the other frequency.
Statistics for the survey are shown in Table~\ref{sourcestats}.

\begin{deluxetable}{cccc}
\tabletypesize{\scriptsize}
\setlength{\tabcolsep}{0.06in}
\tablecolumns{9}
\tablewidth{0pt}
\tablecaption{Source Statistics}
\tablehead{Type & 100 GHz & 150 GHz & Freq Matched}
\startdata
$I$\hfill\hfill             & 289 & 455 & 239 \\
$P$\hfill\hfill             & 3   & 3   & 2   \\
$I$/$P$ Matched\hfill\hfill & 2   & 2   & 1   \\
\enddata
\label{sourcestats}
\end{deluxetable}

In total, 289 (455) sources are detected in \I\ at 100 (150) GHz, of 
which 239 are spatially matched between frequency bands, resulting in 
505 unique sources in total intensity.
Position, major/minor axes, flux, and spectral index for each source in 
\I\ are given in Table~\ref{tab:srccat}. 
Four sources are detected in \it{P}\rm, of which two are matched spatially 
across bands; three of these polarized sources have matching counterparts 
in at least one frequency band in \I. 
Properties of these sources are presented in Table~\ref{tab:polsrccat}.

Simulations are used to determine survey completeness and purity, and accuracy 
of recovered source parameters.
Four different types of simulations (labelled Sim1--Sim4) are used to calculate these 
quantities and how they are affected by features particular to a galactic plane survey, 
specifically the influence of the diffuse background, and the effect of an anisotropic
distribution of spatially clustered point sources.
A detailed description of the simulations is presented in Appendix~\ref{app:sims}.

The 90\% survey completeness in total intensity, $C_{I,90}$, is determined from Sim4 
(the most realistic simulation used, incorporating both detector, atmospheric noise, 
spatially clustered point sources and a model for the diffuse background); we find 
 5.9 and 2.9 Jy at 100 and \fhigh\ respectively.
These do not change significantly with the inclusion/exclusion of a diffuse
background, indicating that the background removal strategy described in Section~\ref{sec:srcextraction}
is effective.

In polarization, the 90\% completeness $C_{P,90}=20.3~\mathrm{Jy}$ at \flow, and 1.1 Jy
at \fhigh.
The large difference is likely due to confusion when projecting source emission of 
randomly oriented polarization angle along the line of sight, an effect reduced at
\fhigh\ on account of the smaller beam size; randomly distributed sources give a 
90\% confusion limit of $\sim1~\mathrm{Jy}$ at both frequencies 
(Appendix~\ref{subsec:completeness}).
Similarly to total intensity, the completeness limit in polarization is not strongly 
dependent on the presence of diffuse galactic emission.

At the signal-to-noise extraction threshold of 5 in \I, the survey is $90-100\%$ pure
at both frequencies, while the purity in polarization is very similar, but at an extraction
signal-to-noise of 3; the spread in purity arises due to the different types of simulation.

The QUaD catalog is matched to the IRAS Point Source Catalog (IRAS-PSC) and the PMN 
catalog~\citep{PMN1993a} using a search radius of $0.1'$.
Of the sources detected at 100 (150) GHz, 97\% (87\%) have IRAS-PSC counterparts;
this discrepency could be due to the similar resolution of IRAS $100\mu$m and QUaD
 \flow, while sources in close proximity may be resolved at \fhigh\ due to the higher 
angular resolution.
Three sources without IRAS-PSC counterpart have associations
with PMN sources.
That such a large fraction of sources have IR counterparts indicates that cores 
located inside the detected clumps are past the prestellar phase and have thermally radiating 
dust envelopes. 
This result might be expected; since the QUaD frequency bands probe the dust emission
well away from the core SED peak, we are unlikely to detect the prestellar or starless
cores which consist solely of very cold molecular gas and have no internal source of
luminosity.

The QUaD catalog can provide constraints on the continuum spectra of 
each source.
However, source fluxes at \flow\ should be interpreted carefully due to the possible 
contribution of free-free emission at this frequency.
For QUaD sources with an \it{IRAS}\rm\ counterpart, \it{IRAS}\rm\ far 
infrared (FIR) fluxes can be used to identify ultracompact \HII\ (UC\HII) regions 
using the Wood-Churchwell~\citep[WC; ][]{wood1989} criterion:
UC\HII\ regions are ionized by O stars, which have very similar flux density 
distributions from object-to-object~\citep{wood1989b} --- the distribution of 
UC\HII\ sources in the FIR color-color plane should therefore be tightly 
restricted, an observation which is the basis of the WC criteria of 
$\log(F_{60}/F_{12})\geq1.30$ and $\log(F_{25}/F_{12})\geq0.57$, where 
$F_{\lambda}$ is the wavelength of the \it{IRAS}\rm\ band in microns.
The QUaD catalog includes a field indicating whether or not each source satisfies 
the WC criteria; if free-free emission is important, the shape of the flux density
distribution will be distorted away from that expected of a UC\HII\ region and 
indicate significant free-free emission.
However, since we cannot assume that all sources in the QUaD catalog are UC\HII\ regions,
we prefer to use the WC criteria simply to indicate whether or not a QUaD source is a UC\HII\
region.
Of the 505 unique sources in the \I\ catalog, 182 satisfy the WC criterion, or 36\% of all
sources detected by QUaD are UC\HII\ regions.

\subsection{Catalog Field Description}
\label{subsec:catfield}

The fields present in the total intensity source catalog (Table~\ref{tab:srccat}) 
are designated as follows, with major and minor axes, intensities and uncertainties 
tabulated for each frequency band.
If a source was spatially matched between frequency bands, the \flow\ coordinates are
quoted.

\newcounter{step}
\begin{list}
{\bfseries\upshape \arabic{step}:}
  {\usecounter{step}
    \setlength{\labelwidth}{2cm}\setlength{\leftmargin}{1.5cm}
    \setlength{\labelsep}{0.5cm}\setlength{\rightmargin}{1cm}
    \setlength{\parsep}{0.5ex plus0.2ex minus0.1ex}
    \setlength{\itemsep}{0ex plus0.2ex} \slshape}
\item Source ID
\item RA: Source right ascension (J2000) in degrees.
\item dec: Source declination (J2000) in degrees.
\item $\sigma_{maj}$: Major axis in arcmin (\flow).
\item $\sigma_{min}$: Minor axis in arcmin (\flow).
\item $I$: Source flux in Janskys (\flow). 
\item $\delta I$: Source flux uncertainty in Janskys (\flow).
\item $\sigma_{maj}$: Major axis in arcmin (\fhigh).
\item $\sigma_{min}$: Minor axis in arcmin (\fhigh).
\item $I$: Source flux in Janskys (\fhigh).
\item $\delta I$: Source flux uncertainty in Janskys (\fhigh).
\item $\alpha_{I}$: Source spectral index.
\item $\delta\alpha_{I}$: Source spectral index uncertainty.
\item $Alt. name$: Alternative name; matched to either IRAS-PSC or PMN catalog.
\item $WC$: Indicates whether the source satisfies the Wood-Churchwell criteria
for ultracompact \HII\ regions.
\end{list}  

The fields present in the polarized intensity source catalog (Table~\ref{tab:polsrccat}) 
are designated as follows, with total intensity and polarized intensity and angle 
tabulated for both frequencies. 

\newcounter{Pstep}
\begin{list}
{\bfseries\upshape \arabic{Pstep}:}
  {\usecounter{Pstep}
    \setlength{\labelwidth}{2cm}\setlength{\leftmargin}{1.5cm}
    \setlength{\labelsep}{0.5cm}\setlength{\rightmargin}{1cm}
    \setlength{\parsep}{0.5ex plus0.2ex minus0.1ex}
    \setlength{\itemsep}{0ex plus0.2ex} \slshape}
\item Source ID
\item RA: Source right ascension (J2000) in degrees.
\item dec: Source dec (J2000) in degrees.
\item $I$: total intensity source flux in Janskys (\flow).
\item $\delta I$: total intensity source flux uncertainty in Janskys (\flow).
\item $P$: Polarized intensity source flux in Janskys (\flow).
\item $\delta P$: Polarized intensity source flux uncertainty in Janskys (\flow).
\item $\phi$: Source polarization angle in degrees (\flow).
\item $\delta \phi$: Source polarization angle uncertainty in degrees (\flow).
\item $I$: total intensity source flux in Janskys (\fhigh).
\item $\delta I$: total intensity source flux uncertainty in Janskys (\fhigh).
\item $P$: Polarized intensity source flux in Janskys (\fhigh).
\item $\delta P$: Polarized intensity source flux uncertainty in Janskys (\fhigh).
\item $\phi$: Source polarization angle in degrees (\fhigh).
\item $\delta \phi$: Source polarization angle uncertainty in degrees (\fhigh).
\item $\alpha_{I}$: Total intensity source spectral index.
\item $\delta\alpha_{I}$: Total intensity source spectral index uncertainty.
\item $\alpha_{P}$: Polarized intensity source spectral index.
\item $\delta\alpha_{P}$: Polarized intensity source spectral index uncertainty.
\item $Alt. name$: Alternative name; matched to either IRAS-PSC or PMN catalog.
\end{list}  

The typical error in each position coordinate are calculated from the 
distribution of position uncertainties taken over all sources; we find 
$\sigma_{x}=^{+0.6}_{-0.1}~\mathrm{arcmin}$ and $\sigma_{x}=^{+0.4}_{-0.1}~\mathrm{arcmin}$
at 100 and \fhigh\ respectively.
Uncertainties in angular size from fitting each source to an elliptical gaussian
function are $\sim0.2~\mathrm{arcmin}$.
We find typical flux uncertainties of $\sigma_{I}=^{+0.94}_{-0.13}~\mathrm{Jy}$ and
$\sigma_{I}=^{+1.2}_{-0.16}~\mathrm{Jy}$ at 100 and \fhigh\ respectively.

\subsection{Source Distribution with Galactic Latitude}
\label{subsec:bdist}

\begin{figure}[h]
\resizebox{\columnwidth}{!}{\includegraphics{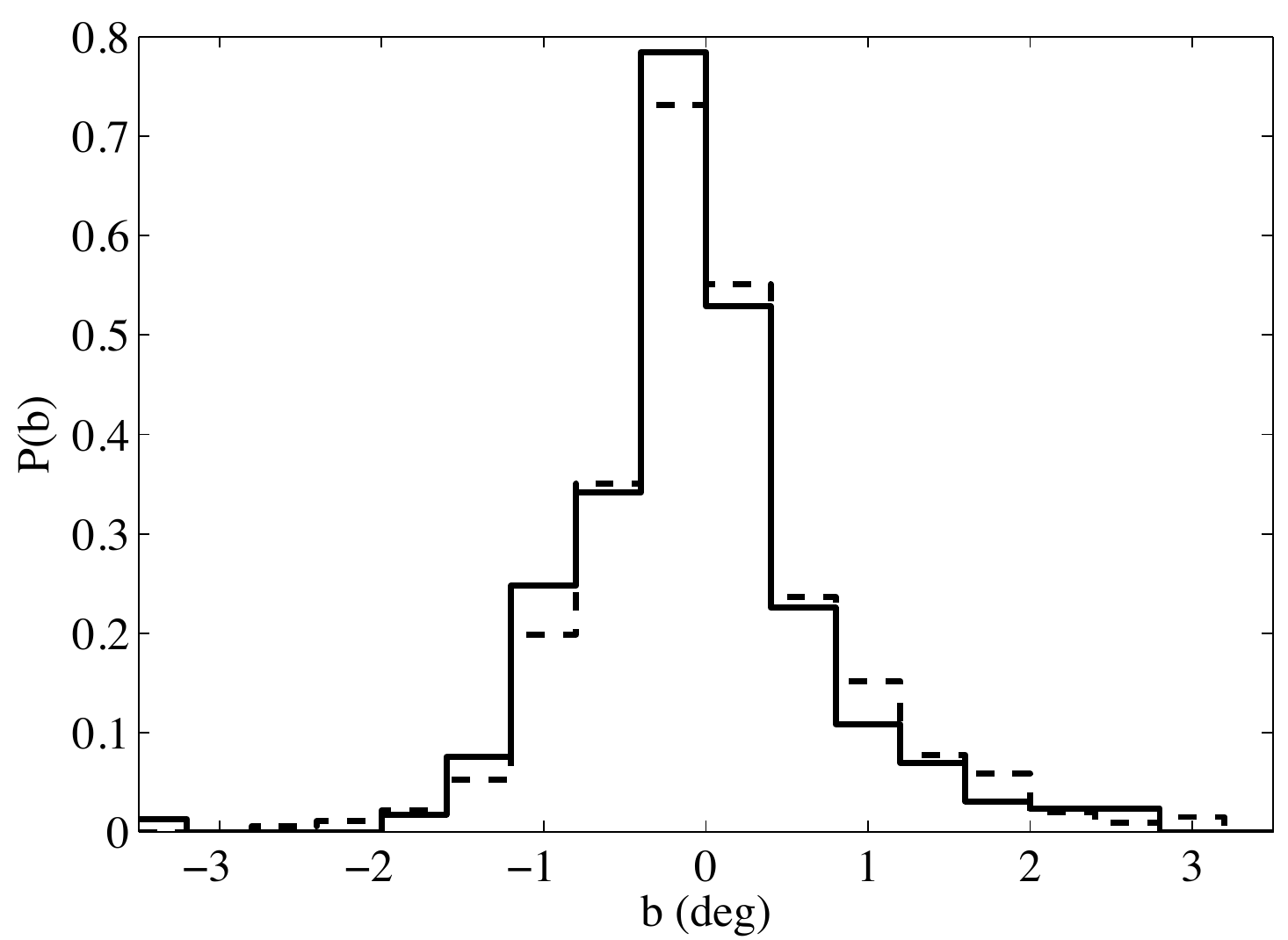}}
\caption{Probability distribution of detected sources in \I\ as a function
of galactic latitude $b$.
The solid (dashed) line is for sources detected at 100 (150) GHz.
At both frequencies, the distribution peaks below $b=0$.}
\label{fig:numvsb}
\end{figure}

Figure~\ref{fig:numvsb} shows the distribution of discrete sources as a function
of galactic latitude $b$, after correction for survey coverage (a smaller range of $b$
is sampled at lower Decl.).
The median of the distribution is $-0.07^{\circ}$ and $-0.04^{\circ}$ at 100 and \fhigh\ respectively.
This negative offset is within one beamwidth of $b=0$ at both frequencies, but supports the results of
other surveys, such as \cite{schuller2009}, who found the peak of the distribution to 
be $-0.09^{\circ}$, at higher ($19.2 ''$) angular resolution.
No obvious explanation for this offset is given, though~\cite{schuller2009} suggest 
the slightly positive galactic latitude of the Sun as a possible cause, or alternatively
the presence of molecular clouds which obscure IR sources; the latter is discussed in the context of 
sources near the galactic center in~\cite{hinz2009}.

\subsection{Source Counts}
\label{subsec:sourcecounts}

\begin{figure}[h]
\resizebox{\columnwidth}{!}{\includegraphics{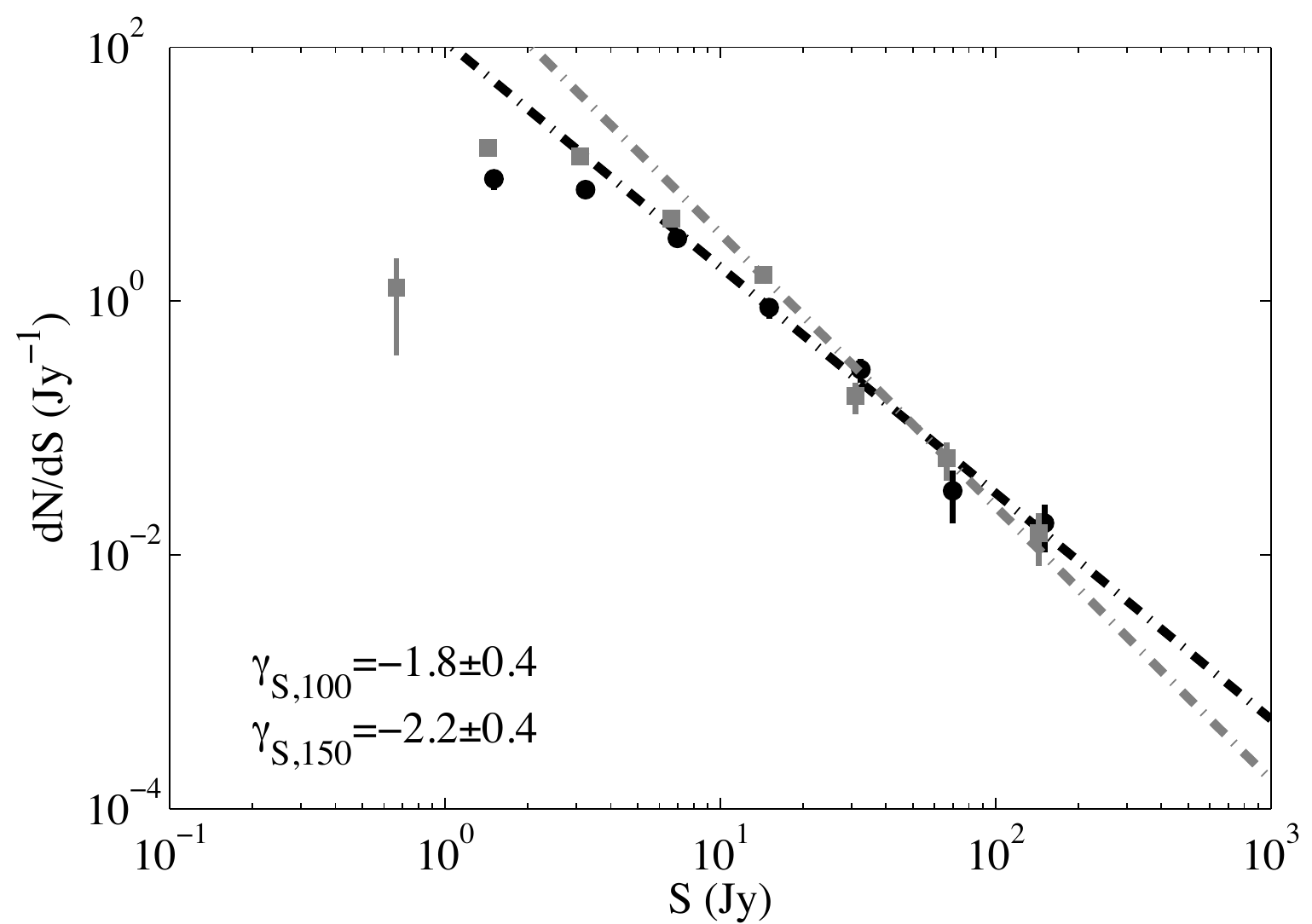}}
\caption{Differential source counts from the survey as a function of flux.
Black is for \flow\ data, gray is \fhigh.
The best-fit slopes are shown in the plots as dot-dashed lines with the same color coding; 
numerical values and uncertainties are shown in the lower left of the plot.}
\label{fig:dNdS}
\end{figure}

Figure~\ref{fig:dNdS} shows the differential source counts of the catalog as a function of 
total intensity flux $S$.
Fitting to a power-law distribution $dN/dS\propto S^{\gamma_{S}}$ in the range $10<S<300$ Jy, 
we find $\gamma_{S,100}=-1.8\pm0.4$ at \flow, and $\gamma_{S,150}=-2.2\pm0.4$ at \fhigh.

If dust dominates the millimeter/sub-mm source emission, their fluxes are proportional 
to the masses of star-forming cores $M$~\citep[e.g.][]{enoch2006}, and
the slope of $dN/dM$ can be used to constrain the slope of the IMF.
Four caveats prevent conversion of QUaD source fluxes to core masses.
First, the resolution required to observe individual cores (as opposed to clumps) is
approximately $30''$, a factor $\sim10$ higher than the QUaD \fhigh\ band.
QUaD sources could in principle contain more than one core, biasing the measurement of
core masses.
Second, since the \flow\ band flux could contain a substantial free-free contribution, 
calculating masses at this frequency is not possible without further information on
the relative contribution of free-free.
This is less of an issue at \fhigh, where the QUaD data should be dominated dust.
Third, the mass conversion also requires a distance estimate to the core which are not
readily available for each source in the catalog.
Fourth, the measurement of the slope may be subject to systematic error due to the presence
of the diffuse background.
Simulations in Appendix~\ref{app:counts} show that if a diffuse background is present 
the slope of $dN/dS$ may not be well-described by a single power-law.
However, this effect depends on the model used for the diffuse emission in the simulations,
namely the amplitude of the background relative to the sources, and the power in diffuse substructure.
Caution is thus advised when interpreting the slope results quoted above, though
Figure~\ref{fig:dNdS} indicates that power-law behavior is observed above 10 Jy,
and thus the contribution of diffuse emission in this flux regime is not important.

Due to these caveats, we caution against overinterpretation of the measured slope of $dN/dS$, 
and refrain from assigning a mass to each source and from converting the slope of 
$dN/dS$ to the slope of the IMF. 

\subsection{Spectral Index Distribution}
\label{subsec:dataspecdist}

\begin{figure}[h]
\resizebox{\columnwidth}{!}{\includegraphics{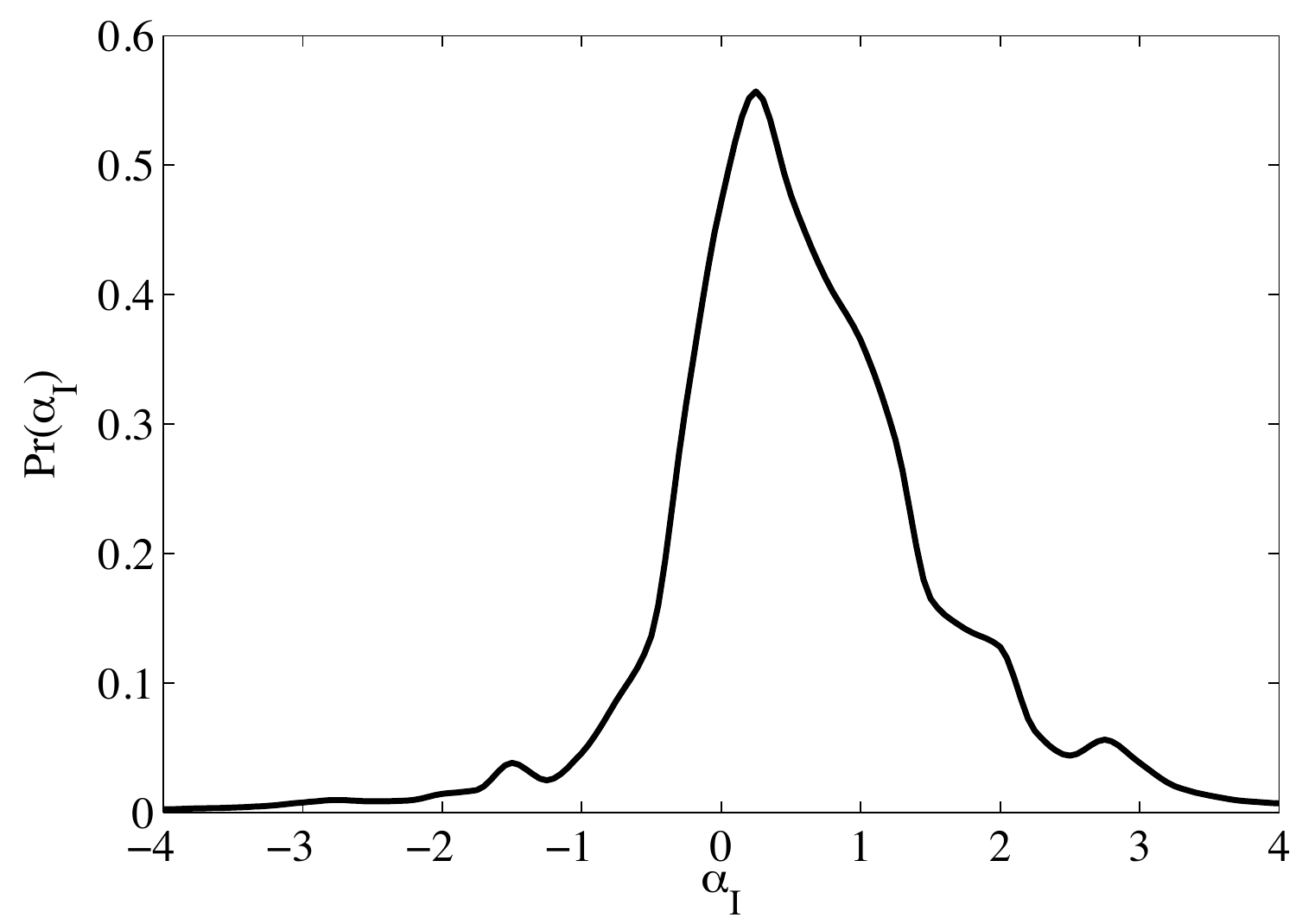}}
\caption{Source total intensity spectral index distribution $Pr(\alpha_{I})$.}
\label{fig:dataspecdist}
\end{figure}

The spectral index distribution in total intensity, $Pr(\alpha_{I})$, is 
computed following~\cite{muchovej2010}.
For each source $j$, the spectral index probability distribution $Pr_{j}(\alpha_{I})$
is calculated by generating flux distributions at each frequency from the 
central value and noise distributions, and then combining the flux distributions.
The spectral index distribution for the sample is then the normalized sum of the
$Pr_{j}(\alpha_{I})$, i.e.
\begin{equation}
\label{eq:alphapdf}
Pr(\alpha_{I})=\frac{\sum_{j}Pr_{j}(\alpha_{I})}{\int\sum_{j}Pr_{j}(\alpha_{I})}.
\end{equation}
\noindent Figure~\ref{fig:dataspecdist} shows $Pr(\alpha_{I})$ for sources 
matched between the QUaD bands.

The spectral index of the sources are somewhat flatter than those found
at higher frequencies~\citep[e.g.][]{desert08}, peaking at $\alpha_{I}\sim0.25$; 
this could be due to the contribution from other emission components at \flow, 
raising the flux at this frequency above that expected from dust alone and therefore 
flattening the spectral index.
Simulations indicate that the spectral index distribution can be slightly skewed towards 
larger $\alpha_{I}$ by background contamination and source confusion (see 
Appendix~\ref{app:specinddist_recovery}); the center of the distribution shifts by 
$\sim0.2$ in $\alpha_{I}$.
Finally, there is evidence of unaccounted emission processes at \flow\ in the QUaD
data (see Map Paper), which would also shift the spectral index 
distribution to lower values; detected sources are generally not faint enough that flux 
boosting is important.

Our analysis does not account for flux boosting due to noise on account of the larger
(systematic) effect of contamination due to the diffuse background and map filtering effects.
Rigorous Bayesian methods to determine the spectral index distribution of 
sources exist in the literature~\citep[e.g.][]{crawford2010,vieira2009},
but do not account for the effect of an unknown background, which is the largest
contaminant to source fluxes in the galaxy as demonstrated in Appendix~\ref{app:srcrecovery}.
We therefore do not pursue such an approach; the increased frequency coverage 
of current-generation satellite experiments such as Planck and Herschel 
may allow an improved treatment of the diffuse background, enhancing the extraction
of discrete galactic sources and their spectral indices.

\subsection{Source Clustering}
\label{subsec:sourceclustering}

Figure~\ref{fig:fdmap} demonstrates that source locations in the QUaD survey are highly 
correlated.
To quantify source clustering, we construct the two-point angular correlation function $w(\theta)$,
defined as the excess probability of finding a source within angle $\theta\pm\Delta\theta$ of another source, 
$H_{d}(\theta)$, compared to the same probability in a distribution of sources with random spatial 
positions, $H_{r}(\theta)$:
\begin{equation}
w(\theta)=\frac{H_{d}(\theta)}{H_{r}(\theta)} -1.
\label{eq:angcorr}
\end{equation}
\noindent ~\cite{enoch2006} model $w$ as a power law in units of projected physical 
separation $r$, $w(r)\propto r^{\gamma_{r}}$, and use the slope as a method of comparing 
the spatial properties of cores in different molecular 
clouds; the authors suggest that different slopes may provide insight into the processes dominating
core fragmentation.
In the QUaD survey, we compute the correlation function in angular units $w(\theta)\propto \theta^{\gamma_{\theta}}$,
 by constructing $H_{d}(\theta)$ from the data, and $H_{r}(\theta)$ from one realization of Sim1
(a simulation with sources distributed randomly over the QUaD survey).
Differing survey areas at each frequency and variations in survey sensitivity are then
accounted for.

Any survey over a large range of galactic latitude faces the problem that the distribution
 of sources is anisotropic, with $w(\theta)$ poorly defined at large galactic latitudes due to
the shape of the galaxy projected on the sky.
In addition, at much smaller separations $w(\theta)$ is not well reconstructed due to the 
large probability that a neighbouring source has a low flux, assuming a power law source 
count $dN/dS\propto S^{\gamma_{S}}$, with $\gamma_{S}<0$.
Therefore, though a bright source will be detected in the survey, its fainter neighbour
is likely to lie below the noise or confusion limit, preventing accurate reconstruction of $w(\theta)$ at small
$\theta$.
Simulations indicate that $w(\theta)$ is well-recovered in the range $0.4^{\circ}<\theta<2^{\circ}$,
(see Appendix~\ref{app:corrfunc}) and therefore these limits are used to fit a power-law to the 
correlation function of the data.
Only sources from the fourth quadrant are used, because the gap in survey coverage between the
third and fourth quadrants introduces artefacts into $w$; since sources in the third quadrant 
account for $<20\%$ of all sources at each frequency, the calculated slope is not
affected by removal of these sources.

\begin{figure}[h]
\resizebox{\columnwidth}{!}{\includegraphics{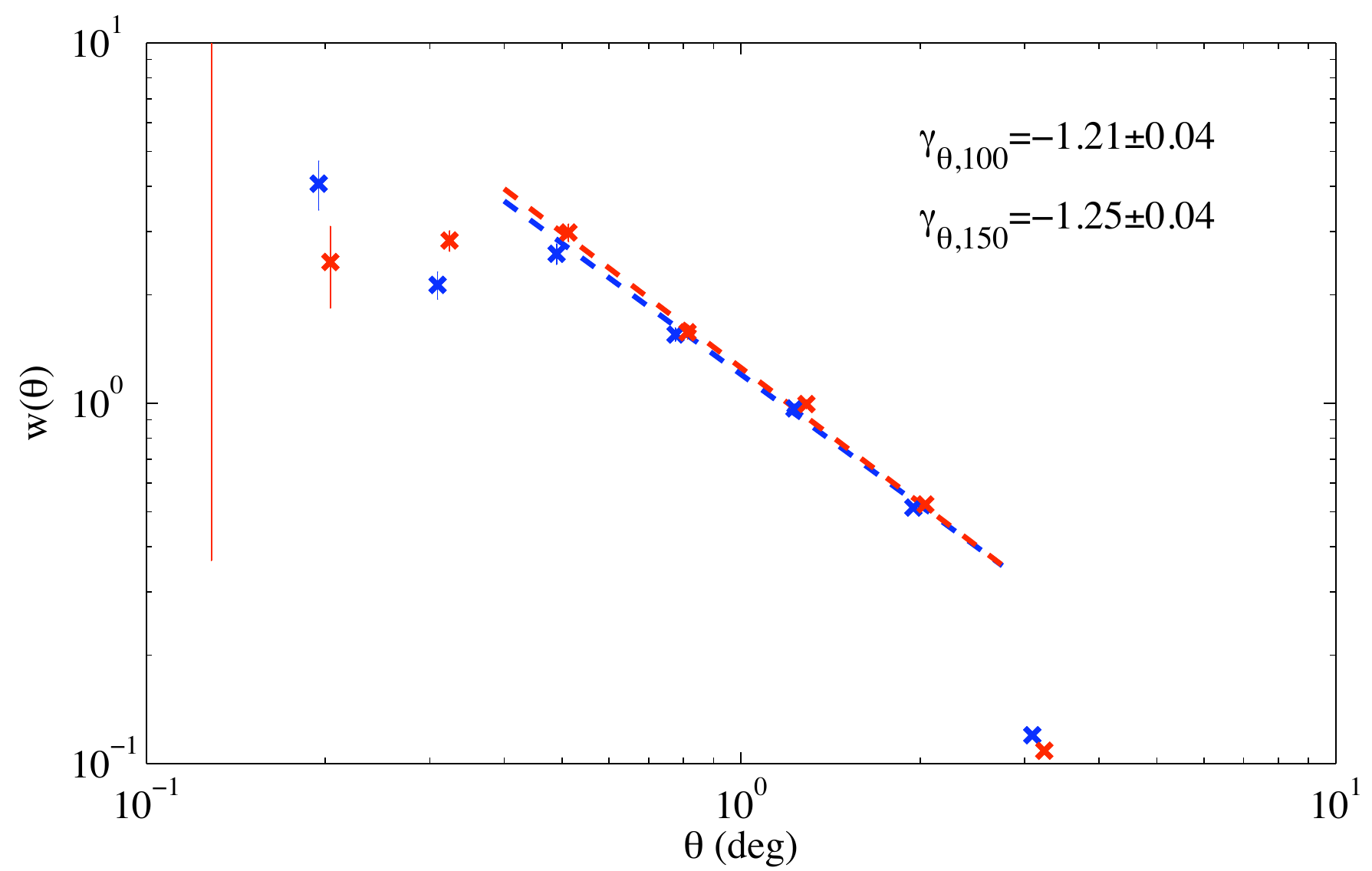}}
\caption{Angular correlation function $w\left(\theta\right)$ for sources in QUaD survey;
Blue is for \flow\ data, red is \fhigh.
Only fourth quadrant data is used because the gap in survey coverage between the third
and fourth quadrants introduces a discontinuous range of source galacic longitudes.
Since the number of sources in the third quadrant is only a small fraction of the total
survey, the effect on the analysis is small.
}
\label{fig:corrfunc}
\end{figure}

Figure~\ref{fig:corrfunc} shows the results. 
We find a power-law slope of $\gamma_{\theta,100}=-1.21\pm0.04$ and $\gamma_{\theta,150}=-1.25\pm0.04$,
consistent with the value found by~\cite{enoch2006} for the Bolocam observations of the Perseus 
molecular cloud, $w(r)\propto r^{-1.25}$.
Since a single distance is assumed to Perseus, there is a one-one mapping between $r$ and $\theta$,
implying that the same correlation function slope applies to sources on large and small angular 
scales (QUaD and Bolocam respectively).
However, it should be noted that due to differing resolutions, Bolocam and QUaD measure the 
angular correlation function of different types of source; cores in the case of Bolocam, and 
clumps in the case of QUaD.
Therefore, while the correlation function slopes are consistent, it is not clear that the 
correlation functions measured by each experiment are directly related, and thus caution is advised when 
comparing these results.

\subsection{Polarized Sources}
\label{subsec:polsources}

\begin{figure}[h]
\resizebox{\columnwidth}{!}{
\includegraphics{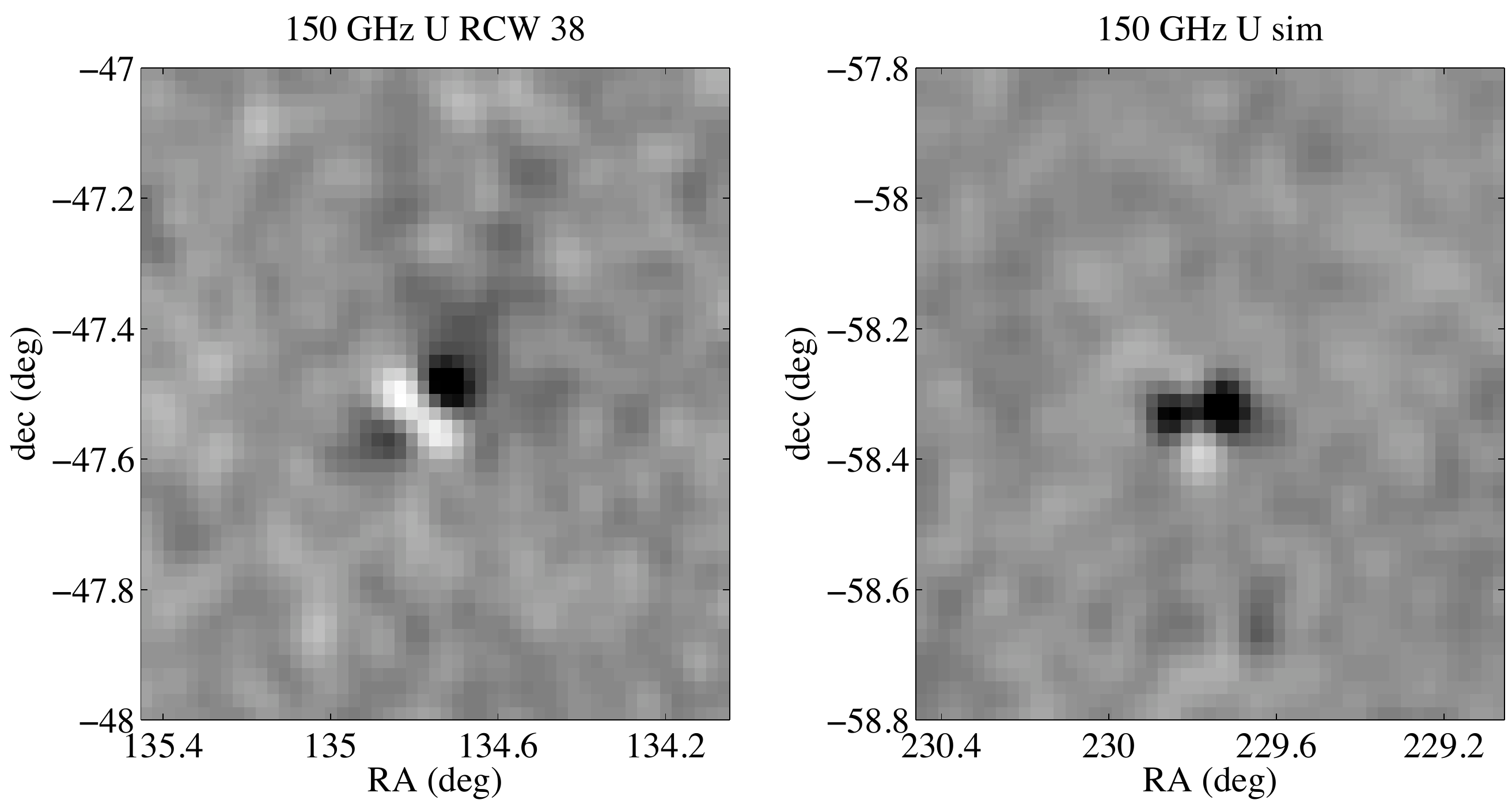}   
}
\caption{Illustration of optical effect for low ($<1\%$) fractional
polarization sources.
Left is \fhigh\ \U\ map centred on RCW 38; right is a \fhigh\ \U\ map of a simulated
source showing a similar optical effect.
The colorscale on each plot is the same.}
\label{fig:opticalfx}
\end{figure}

Polarized sources are of particular interest due to their scarcity and the fact that
they offer a means to probe small-scale magnetic field structure in the galaxy.
Maps of detected polarized sources are examined visually in order to reject
beam-scale optical effects (such beam offsets and/or differing beam ellipticities between 
two PSBs within a feed).
These can cause spurious source detection at fractional polarization ($\sim 1\%$ or less)
 --- see Instrument Paper for further details.
Figure~\ref{fig:opticalfx} shows the \fhigh\ \U\ map of RCW 38 and a simulated point source 
of low fractional polarization.
The `quadrupole' polarization pattern is observed in the two cases, implying that 
the apparent polarization of RCW 38 is an instrumental effect rather than real polarized signal.
Polarized sources exhibiting such a pattern are visually rejected from the catalog.

\begin{figure*}[ht]
\resizebox{\textwidth}{!}{
\includegraphics{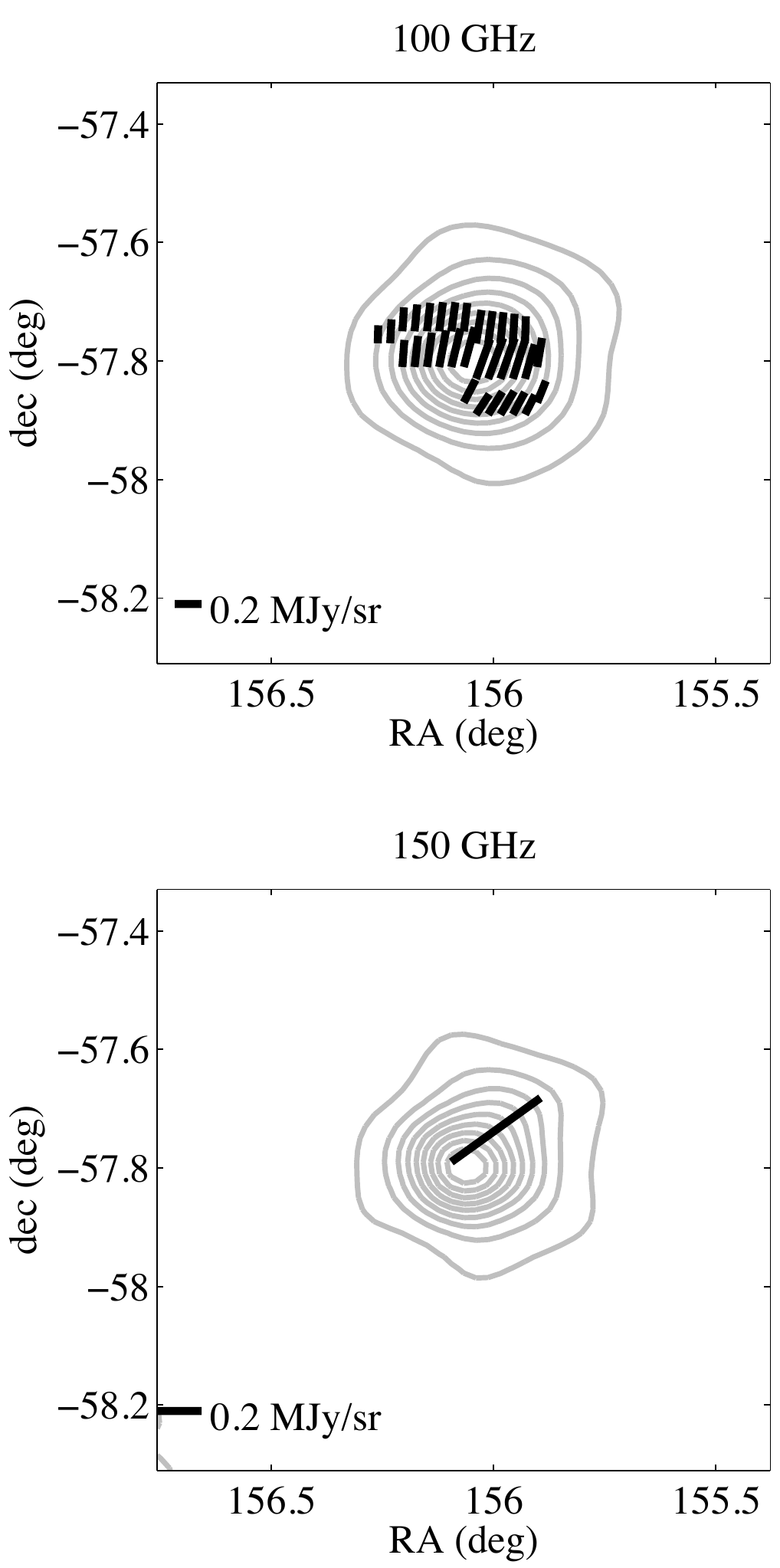}   
\includegraphics{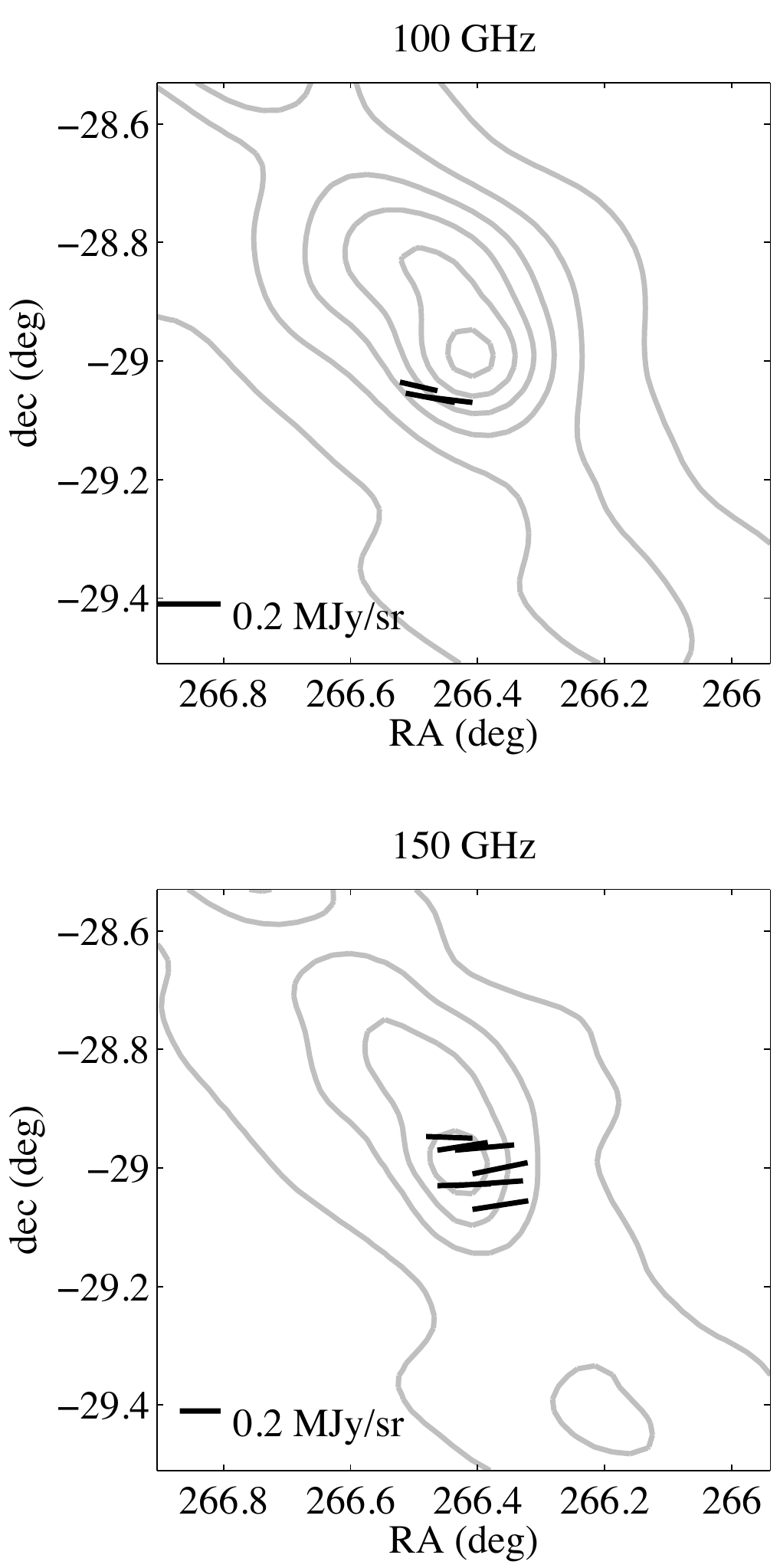}   
\includegraphics{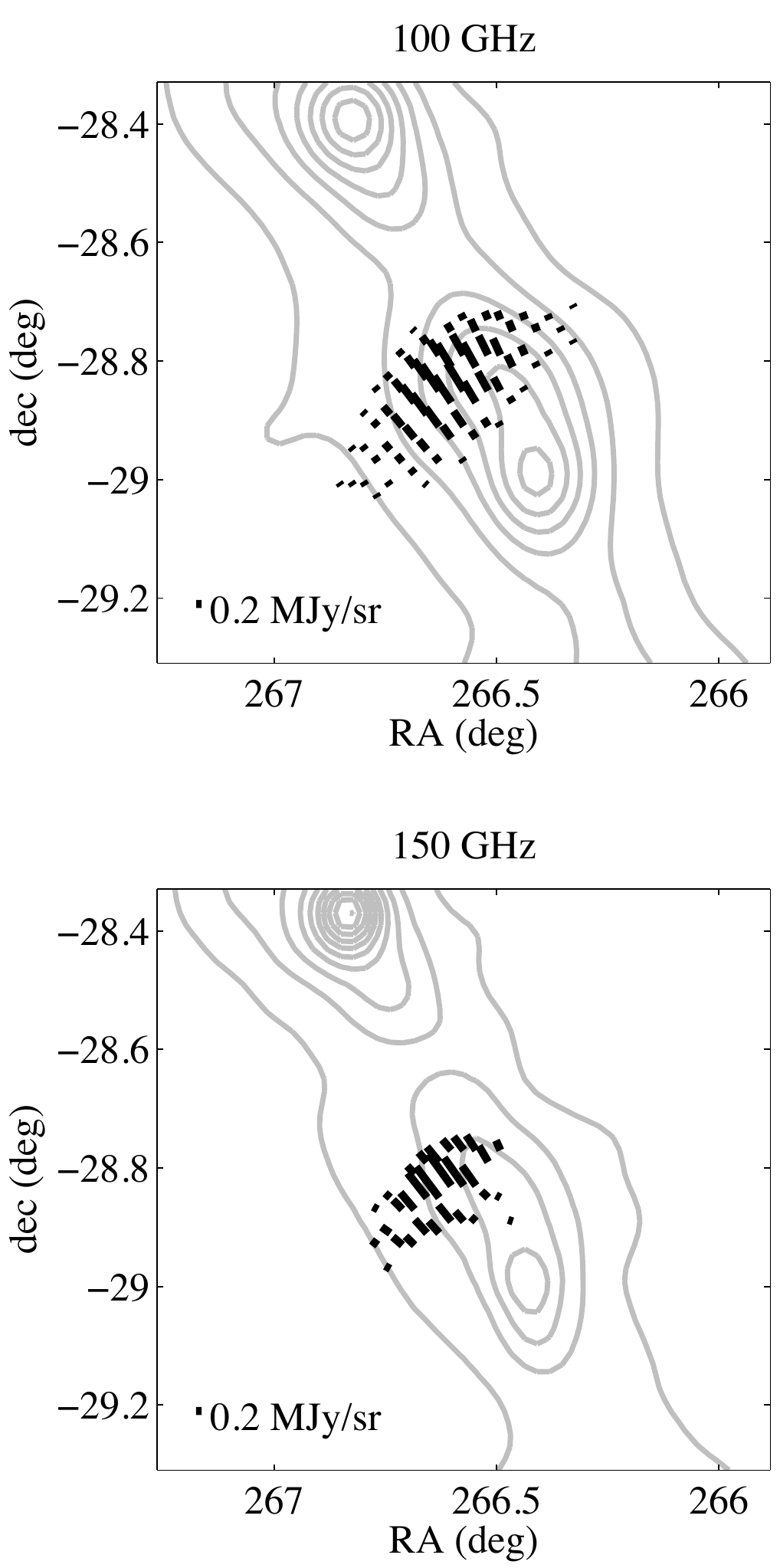}   
\includegraphics{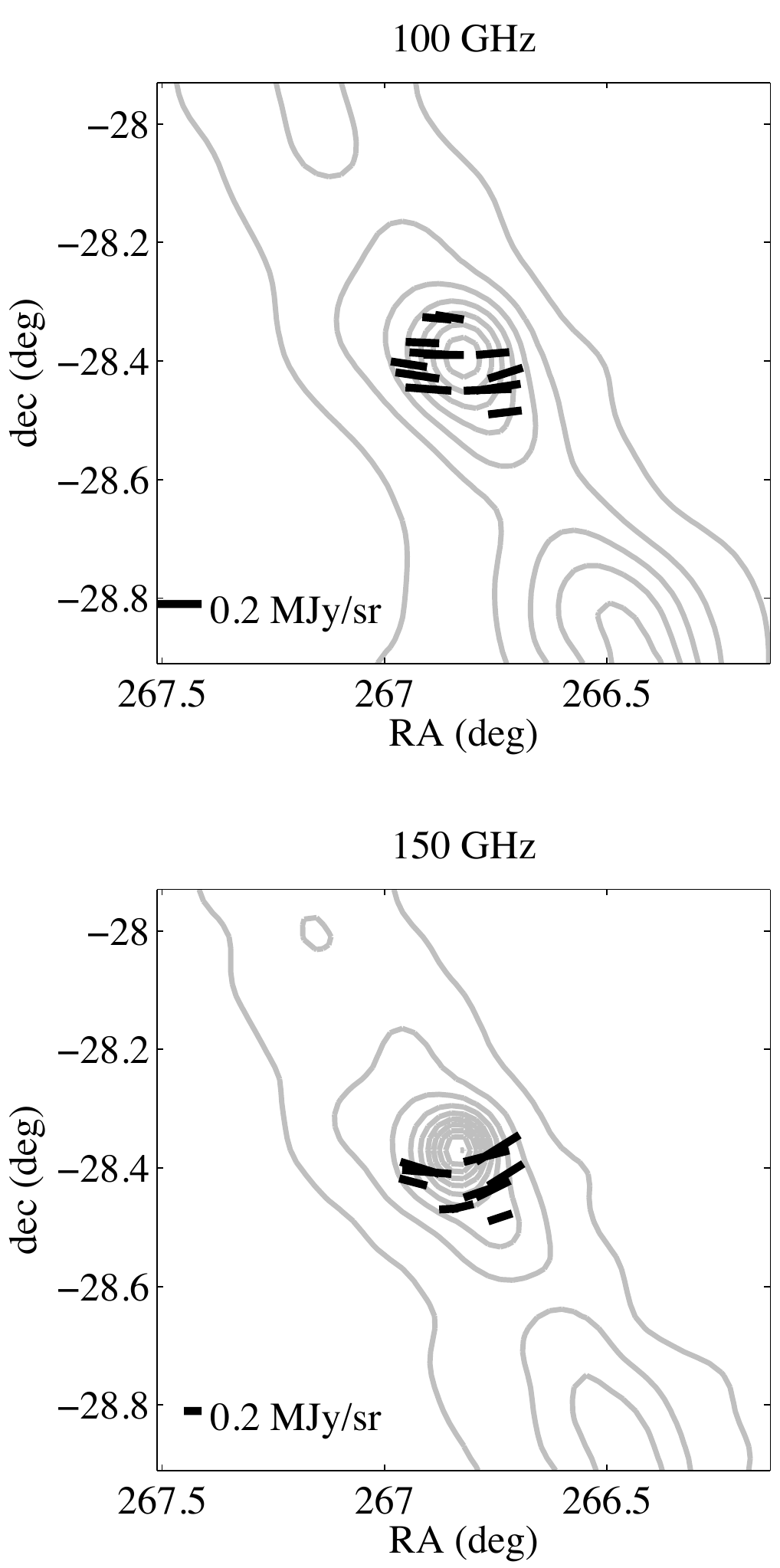}   
}
\caption{Total intensity images with polarization vectors overlaid on each source detected
in polarized intensity --- source fluxes, spectral indices and alternative identifications are
presented in Table~\ref{tab:polsrccat}.
Top row is \flow, bottom row is \fhigh, only polarization vectors with signal-to-noise $>5$ are plotted.
From left to right: 284.33-0.36 (RCW 49), 359.93-0.06 (Sagittarius A*), 0.18-0.06 (Galactic Center Arc), 
and 0.63-0.05 (Sagittarius B2).
For 284.33-0.36, the contours run from 1.6 to 16.1 MJy/sr in steps of 1.6 MJy/sr at \flow, and from
2.3 to 23.3 MJy/sr in steps of 2.3 MJy/sr at \fhigh.
For the remaining sources, the contours run from 2 to 20 MJy/sr in steps of 2 MJy/sr at \flow, and
from 5 to 50 MJy/sr in steps of 5 MJy/sr at \fhigh.
}
\label{fig:polsrc}
\end{figure*}

Properties of remaining polarized sources in the QUaD survey are presented in 
Table~\ref{tab:polsrccat}, with images of each in total and polarized intensity in
Figure~\ref{fig:polsrc}.
Below we discuss each source in more detail.

\subsubsection{284.33-0.36: RCW 49}
\label{subsec:pol1} 

RCW 49 is a bright \HII\ region covering $90'\times70'$, which is being ionized by the rich, 
compact star cluster Westerlund 2~\cite[e.g.][]{furukawa2009}.
Numerous total intensity observations exist from the radio to X-ray wavebands; however, existing
studies of this source in polarization near the QUaD bands have been restricted to~\cite{dickinson2007},
who observed several southern \HII\ regions with the CBI telescope at 31~GHz.
Their measurements of RCW 49, at 6.78 arcmin angular resolution, provide an upper limit on
the 31~GHz polarization fraction of 0.24\%, limited by instrumental leakage from Stokes $I$
to $Q$ and $U$.
Below 31~GHz, the RCW 49 emission is dominated by free-free, as indicated by the total intensity
spectral index $\alpha_{I,RCW 49}=-0.220\pm0.074$ between 2.7 and 15~GHz~\citep{dickinson2007}.
The QUaD total intensity counterpart to RCW 49, 284.33-0.36, also indicates a flat spectral
index between 100 and~\fhigh\ of $\alpha_{I,RCW 49,QUaD}=-0.08\pm-0.01$.
Polarized emission is detected in the QUaD data at \flow\, with a polarization fraction of $0.019\pm0.0076$ and
the polarization vectors aligned predominantly east-west (see Figure~\ref{fig:polsrc}).
The absence of detected polarization at \fhigh\ indicates the emission may
not be thermal in nature; this idea is supported by the total intensity spectral index, though
this measurement is likely biased flat by the presence of free-free emission. 
While free-free is not intrinsically polarized, it may cause polarization by Thomson scattering 
at the edges of the \HII\ region, resulting in tangentially polarized radiation at the cloud 
edges.
Since the QUaD \flow\ polarization vectors are largely aligned over the source area, free-free
polarization at the cloud edges can be ruled out.
Synchrotron radiation is a further possibility, but unlikely given the physical nature of the 
source. 
It is therefore possible that instrumental effects other than those illustrated in 
Figure~\ref{fig:opticalfx} are present.

\subsubsection{359.93-0.06: Sagittarius A*}
\label{subsec:pol2} 

High frequency polarized observations of Sagittarius A* (Sgr A*, detected in the QUaD survey as
359.93-0.06) are important due to the constraints they provide on relativistic jets
and accretion processes in black holes~\citep[e.g.][]{agol2000,quataert2000,melia2001}.
The rotation measure (RM), which is proportional to the electron density and magnetic field component 
integrated along the line of sight, can be determined from multi-frequency observations of the 
linear polarization fraction --- measurements of the RM constrain the mass accretion rate 
of the black hole and thus rule out certain classes of accretion model.
Observations of linear polarization in Sgr A*, first detected with SCUBA by~\cite{aitken2000}, have 
been studied with interferometric imaging at frequencies above $\sim100~\mathrm{GHz}$, 
typically with resolution $<10''$ or better~\citep{macquart2006,marrone2007}.
At 3.5mm,~\cite{macquart2006} find a fractional linear polarization of $2.1\%\pm0.1\%$, while larger
values are found at higher frequencies ($\sim5\%$ and $\sim9\%$ at 230 and 340~GHz respectively,
~\citet{marrone2007}).
~\cite{aitken2000} measured an observed polarization fraction of $2.9\%\pm0.3\%$ at 2mm with $33.5''$
resolution.

In the QUaD data, Sgr A* is detected in polarization at \fhigh\ only (see Figure~\ref{fig:polsrc} and 
Table~\ref{tab:polsrccat}), but the polarized flux is unconstrained; it appears that the 
polarization fraction is $<1\%$ at \fhigh, considerably less than that found by interferometric 
instruments at comparable frequencies.
This result is not surprising, given the factor $\sim20$ lower resolution compared to BIMA
at 3.5mm~\citep{macquart2006}, and a factor $\sim6$ lower than SCUBA at 2mm~\citep{aitken2000}.
At $3.5'$ angular resolution at \fhigh, the polarized emission from Sgr A* is smeared out and 
the polarization fraction therefore reduced due to the contribution from the diffuse background or 
unpolarized sources within a QUaD \fhigh\ beam of Sgr A*.
Though QUaD observations span four months and could constrain the variability of the polarized 
emission (a useful diagnostic of processes intrinsic to the source), at such low resolution and 
signal-to-noise it is not possible to measure this effect.
The apparent low polarization fraction of this source may result in false detection due to an 
optical effect in the telescope, similarly to 284.33-0.36, and should therefore be treated with caution.

\subsubsection{0.18-0.06: The Galactic Center Arc}
\label{subsec:polarc} 

0.18-0.06 is clearly associated with G0.2-0.0, also known as the Galactic Center Arc.
This source has an extent of $\sim25~\mathrm{arcmin}$ along its long axis perpendicular
 to the galactic plane, is approximately symmetric with respect to the galactic equator, and
is among the brighter sources close to the galactic center~\cite[e.g.][]{altenhoff1979,lis1994}.
High resolution observations~\citep[e.g.][]{yusef2004} indicate this filamentary structure contains
the largest concentration of nonthermal radio filaments in the galaxy. 
Though this object appears extended and unassociated with any discrete source in the QUaD \I\ data (see
Figure~\ref{fig:polsrc} and Table~\ref{tab:polsrccat}), the radio filaments cross three 
bright \HII\ regions unresolved by QUaD~\citep[e.g.][]{yusef1986,reich2000}: G0.16-0.15, G0.18-0.04 and G0.1+0.08.
Using the Green Bank Telescope,~\cite{law2008} determined that this radio arc has a non-thermal
spectrum of $-0.54\pm0.09$ between 4.85 and 8.5~GHz (resolution 2.5 and 1 arcminute respectively), 
in support of the idea that the emission is not from cold dust, but rather from either monoenergetic 
electrons or an electron distribution with a low energy cutoff~\citep{reich2000}.
Observations at 4.8~GHz using the VLA~\citep{yusef1986} demonstrate that while thermal emission
dominates G0.18-0.04 and G0.1+0.08, the low-frequency nonthermal polarized emission in the arc is
 primarly due to the \HII\ region G0.16-0.15; the high ($\sim30\%$) fractional polarization observed 
provides further evidence that the polarized emission is due to synchrotron.

We find a peak polarization fraction of $\sim10\%$ at both 100 and~\fhigh, and a 
polarized spectral index of $-1.04\pm0.17$, indicative of a source dominated by synchrotron emission.
Since we do not detect a discrete total intensity source associated with the polarized arc,
we estimate the polarization fraction from the raw \I\ images and the polarized flux
measured by the source extraction algorithm.
Due to its spatial extent above the beam scale in the QUaD survey, it is doubtful that the radio 
arc's polarized emission at these frequencies can be solely attributed to G0.16-0.15, which
would be likely be unresolved.
The high level of uniformity of the polarization seen in Figure~\ref{fig:polsrc} indicates a highly ordered
magnetic field which must exist over the full extent of the arc, rather than localized to a single 
\HII\ source.
The polarized vectors are largely aligned parallel to the plane of the galaxy, almost perpendicular
to the polarization from diffuse emission (see Map Paper), indicating a strong local deviation 
from the galactic magnetic field.

\subsubsection{0.63-0.05: Sagittarius B2}
\label{subsec:pol4} 

The polarization of Sagittarius B2 (Sgr B2) has been well-studied in the radio, sub-millimeter and 
far-infrared~\citep[e.g.][]{greaves1995,dowell1997,novak1997,dowell1998,jones2010}; the QUaD data fill
in the millimeter portion of the spectrum of this giant molecular cloud.
Observations of linearly polarized emission from magnetically aligned dust grains in such clouds
can be used to determine the orientation of the local magnetic field.
In the sub-millimeter, where the dust is optically thin, the polarization is due to emission
of the dust grains preferentially along their long axis; \cite{greaves1995} observed Sg B2 at 
800$\mathrm{\mu m}$ and $30''$ resolution, finding polarization fractions in the range $0.8\%-2.6\%$
with the polarization orientation approximately north-south.
Polarization observations at far-infrared wavelengths include both emission and absorption effects.
The former is due to dust grain emission similarly probed by sub-millimeter observations and results
in polarization vectors aligned perpendicular to the local magnetic field direction, while the latter
is caused by absorption from cold, magnetically aligned dust grains in regions of high optical depth, with
corresponding polarization vectors aligned parallel to the magnetic field.
\cite{novak1997} demonstrate that at $115~\mathrm{\mu m}$ with a $35''$ beam, Sgr B2 is resolved
 into `core' and `envelope' regions, with polarization in the core and envelope dominated by absorption
 and emission respectively.
In their study, the envelope fractional polarization ranges from 2--4\%.
At $350~\mathrm{\mu m}$, the effects of absorption are diminished due to the increasing contribution
of dust emission and the decreasing optical depth; \cite{dowell1998} show this observationally,
with the core polarization ($\sim1\%$ fractional polarization) smaller than that in the 
envelope ($\sim2.8\%$) at $20''$ resolution.

Based on the above considerations, in the QUaD bands the polarization of Sgr B2 
should be dominated by dust grain emission processes.
The best measurement of polarization fraction is at \fhigh, where we find a value of $1\%\pm0.2\%$.
It is not surprising that the \fhigh\ polarization fraction is smaller than at $350~\mathrm{\mu m}$.
Since the QUaD beam at this frequency is a factor $3.5\times60 / 20\simeq11$ larger, the core and 
envelope are not resolved into separate components, and thus the higher envelope polarization fraction
is biased low by the less strongly polarized core region.
The east-west orientation of the \fhigh\ QUaD polarization vectors (see Figure~\ref{fig:polsrc}), 
similar to those at $350~\mathrm{\mu m}$~\citep{dowell1998}, further indicates that we are observing 
polarized dust emission from the envelope.

\subsubsection{An Extended Polarized Source: Further Filtering Considerations} 
\label{subsec:polfilt} 

For the sake of simplicity, the same aggressive background filtering (small $\sigma_{bck}$ relative 
to $\sigma_{beam}$) is used in both total and polarized intensity.
This choice was motivated by the bright diffuse emission in total intensity: a larger $\sigma_{bck}$ 
would result in a smaller systematic loss of signal from point sources but an increased contribution
from the background --- see Appendix~\ref{subsec:sigmabck}.
However, this choice may be relaxed in polarized intensity because the polarized diffuse emission 
is relatively faint (polarization fraction $<2\%$, see Map Maper).
The present catalog is therefore not optimized for the detection of polarized sources, and more 
sources could in principle be extracted from the maps.

\begin{figure}[h]
\resizebox{\columnwidth}{!}{
\includegraphics{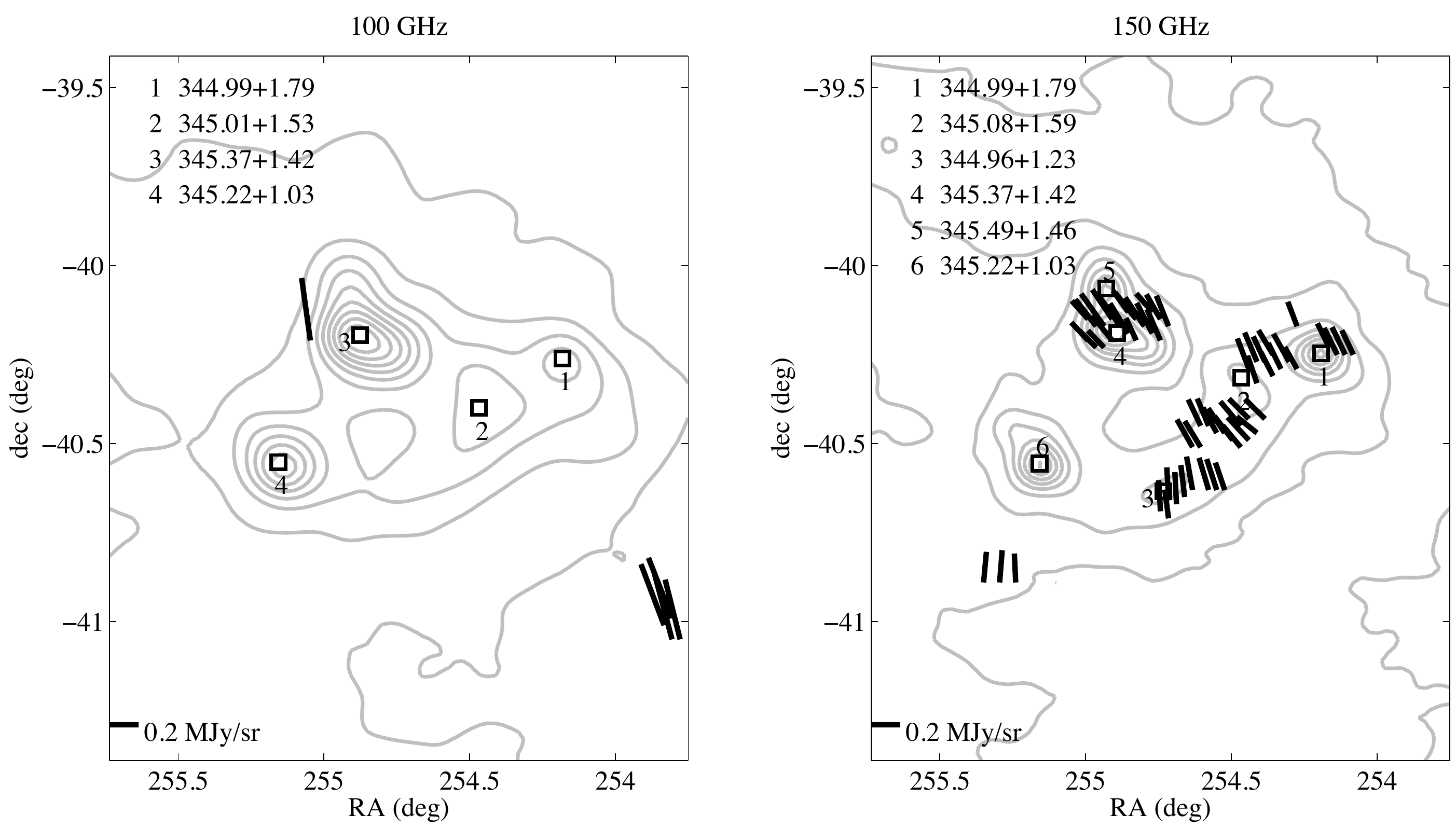}   
}
\caption{Large-scale polarized cloud at \flow\ (left) and \fhigh\ (right) undetected by
source extraction algorithm --- only polarization vectors with signal-to-noise 
$>4$ are plotted (black lines).
At \flow, the contour scale (gray lines) runs from 0.32 to 3.2 MJy/sr in steps of 0.32 
MJy/sr, and from 0.77 to 7.7 MJy/sr in steps of 0.77 MJy/sr at \fhigh.
Black squares indicate the locations of discrete sources detected in the \I\ maps, with
the names of the sources as indicated at upper left of each plot.
}
\label{fig:polcloud}
\end{figure}

As an illustration, Figure~\ref{fig:polcloud} shows total and polarized intensity maps for a visually
identified cloud with appreciable polarization at \fhigh\ (up to 10\% fractional polarization).
Since this source has a relatively large extent in comparision to the beam scale, it is undetected 
by the source extraction algorithm because most of the polarized flux is filtered out in the 
background removal step.

This object is host to several discrete sources, namely 344.99+1.79, 345.01+1.53, 345.37+1.42 
(resolved into 345.38+1.41 and 345.49+1.46 at \fhigh), 345.22+1.03, and 344.96+1.23 (detected at
\fhigh\ only). 
Sources 345.01+1.53 and 345.08+1.59, detected at 100 and \fhigh\ respectively, are cleary associated
with the same source, but have observed centroids separated by more than the $0.02^{\circ}$ required to 
meet the spatial matching criterion.
The polarized emission in this region is only partially correlated with the positions of the
discrete \I\ sources (see Figure~\ref{fig:polcloud}), and the polarization vectors are largely
oriented perpendicular to those from the bulk polarized galactic emission.
This indicates that the diffuse dust is subject to a local magnetic field strong enough to 
overcome that of the galaxy as a whole, despite its small galactic latitude of $\sim1.4^{\circ}$.
Clearly this source is worthy of detailed follow-up study.

\section{Discussion and Conclusions}
\label{sec:conclusions}

We present a catalog of discrete sources extracted from the QUaD galactic plane 
survey, which spans approximately 245-295$^\circ$ and 315-5$^\circ$ in galactic longitude
 $l$ and -4 to +4$^\circ$ in galactic latitude $b$ --- a total of $\sim800$ square degrees
coverage in Stokes \I, \Q, and \U\ at 100 and \fhigh, with resolution 5 and 3.5
arcminutes respectively.
Simulations of a toy model galaxy including spatially clustered point sources 
and diffuse emission indicate a 90\% completeness flux of 5.9 (2.9) Jy at 100 (150) 
GHz in \I, and 20.3 (1.1) Jy in polarization at 100 (150) GHz. 
The high \flow\ completeness in polarization is due to source confusion in the
larger beam, and is dependent on the parameters used for the toy galactic model;
randomly distributed sources yield a completeness of 1.3 Jy 
--- see Appendix~\ref{subsec:completeness} for a full discussion.
At a signal-to-noise threshold of 5 (3) in total (polarized) intensity, 
the catalog is 98\% pure in \I\ at both frequencies, and 97\% (92\%) pure 
in polarization at 100 (150) GHz.
Simulations without a diffuse background are used in the total intensity 
computation because substructure in diffuse emission, detected as discrete sources,
 biases the purity low.
While this could also affect the real catalog, the high percentage of IRAS-PSC 
counterparts to QUaD sources (97\% and 87\% at 100 and \fhigh\ respectively)
indicates that this effect is likely very small.
The polarized diffuse background, with fractional polarization $\sim2\%$ (see
Map Paper), is faint enough that it does not bias the purity of the catalog 
at the signal-to-noise threshold of the survey.
Instrumental effects prevent detection of polarized sources with polarization 
fraction $\sim1\%$ or less. 

In total intensity the catalog contains 505 unique sources, of which 239 are spatially
matched between frequency bands, with 50 (216) detected at 100 (150) GHz alone.

The \I\ flux distributions are well-approximated by a power law over
more than two orders of magnitude above $\sim10$ Jy at both frequencies.
We find power-law slopes of $\gamma_{S,100}=-1.8\pm0.4$ at \flow, and 
$\gamma_{S,150}=-2.2\pm0.4$ at \fhigh; the latter is consistent 
with~\cite{rosolowsky2009}, who find $-2.4\pm0.1$ at 268~GHz
with Bolocam at higher resolution.
The flatter slope at \flow\ may be the result of resolution effects due to the
larger beam at this frequency. 
Simulations indicate that if the diffuse background contributes spurious sources,
as expected the recovered source flux distribution does not accurately follow the underlying 
distribution; however, as discussed above, the high percentage of QUaD sources
spatially matched to IRAS indicates this effect is insignificant.

The spectral index probability distribution of sources in total intensity is 
found to peak at $\alpha\sim0.25$, flatter than expected for sources whose 
emission is dominated by thermal dust.
Simulations indicate the diffuse background does not strongly influence source
spectral indices; the flatness is therefore likely due to free-free emission,
which becomes significant at $\sim$\flow\ and below. 
At this frequency, free-free results in higher fluxes than expected from dust 
alone, shifting the spectral index distribution to lower values.

We explore the clustering of galactic sources by fitting the two-point correlation 
function to a power-law using the \I\ source locations.
Simulations indicate that the underlying correlation function slope can be 
accurately reconstructed in the range $0.4^{\circ}<\theta<2^{\circ}$, with 
$\theta$ the angular separation between a pair of sources.
The correlation function breaks down at larger angular scales because so few 
($<1\%$) of sources are located beyond $|b|=3^{\circ}$.
At galactic latitudes smaller than $0.4^{\circ}$, $w(\theta)$ is not well 
reconstructed because for a power-law $dN/dS\propto S^\gamma_{S}$ with $\gamma_{S}<0$, the survey
does not detect most neighbors of a source bright enough to be included --- one
must extend the search to large angular separations before enough neighbors are
detected for accurate reconstruction.
Fitting to the QUaD \I\ catalog data in the range $0.4^{\circ}<\theta<2^{\circ}$, 
we find power-law slopes of $\gamma_{\theta,100}=-1.21\pm0.04$ and 
$\gamma_{\theta,150}=-1.25\pm0.04$ at 100 and \fhigh\ respectively.
These are consistent with the value found by~\cite{enoch2006} Bolocam 
observations of the Perseus molecular cloud, $w(\theta)\propto \theta^{-1.25}$, 
though the results are not directly comparable on account of the different sources
probed by QUaD (clumps) and Bolocam (cores) due to their differing angular resolution.

97\% (87\%) of the sources detected at 100 (150) GHz have IRAS-PSC counterparts. 
These large fractions indicate that most of the clumps detected in the survey
are past the prestellar phase and have envelopes heated by protostars.
This observation might be expected, given that the QUaD frequency bands lie far
from the spectral peak: Only these sources are bright enough in the Rayleigh-Jeans
portion of the spectrum to be detected in the QUaD survey, unlike prestellar or 
starless sources.
Since the QUaD survey is sensitive to free-free emission as well as dust, 
particulary in the \flow\ band, sources might also be detected if their gas is 
sufficiently ionized to produce free-free but their envelopes are yet to thermalize.
However, the small fraction of \flow\ QUaD sources unmatched to IRAS-PSC imply that almost
all the detected sources at this frequency do have a thermal component.
At \fhigh, the larger unmatched fraction is likely due to single IRAS sources being 
resolved into two sub-clumps by the the higher QUaD resolution at this frequency --- 
only one of these sub-clumps can be spatially matched to the IRAS source.

Of the sources with an IRAS counterpart, 182 satisfy the Wood-Churchwell criteria
for ultracompact \HII\ regions~\citep{wood1989}, providing new spectral constraints
on this class of object.

Four compact polarized sources were detected by the automated source-finding algorithm:
284.33-0.36 (IRAS 10227-5730 or RCW 49), 359.93-0.06 (Sagittarius A*), 0.18-0.06 (IRAS 17431-2846
or Galactic Center Arc), and 0.63-0.05 (IRAS 17440-2823 or Sagittarius B2).
One additional extended source was located `by eye' from the raw \Q\ and \U\ maps; this object
appears host to several discrete total intensity sources, including 344.99+1.79, 345.01+1.53,
345.37+1.42, and 345.22+1.03.
The brightest polarized source is 0.18-0.06, which does not have an obvious discrete 
counterpart in total intensity, but has a polarization fraction of $\sim10\%$ if the diffuse 
background is used as a measure of \I.
It has a polarized flux of $7.91\pm0.33$ ($4.90\pm0.32$) Jy at 100 (150)~GHz,
and a polarized spectral index of $\alpha_{P}=-1.04\pm0.17$, indicating a synchrotron
emission source.
Its detection against a polarized background implies that there is
a strong local deviation from the galactic magnetic field.

Less than 1\% of the sources detected in \I\ have a polarized counterpart. 
If discrete sources do not harbor strong local magnetic fields or shielding, 
dust grains in their envelopes will align with the large-scale galactic field.
The only way to separate diffuse from discrete polarized emission
would then be via morphology (similar to \I) or spectrally, since the orientation of the 
polarization would be similar for diffuse and discrete sources.
Alternatively, the discrete total intensity sources may have fractional polarization $<1\%$,
 as might be expected from a star-forming clump, in which case instrumental effects 
prevented detection of their polarized emission here.
Discrete sources may therefore not be a significant contributor to the low-latitude 
galactic polarized emission.
More sensitive observations (such as from the Planck satellite) will be needed to better 
study the polarization of these sources, and the role of magnetic fields in star-forming regions.

The QUaD catalog may prove useful for a variety of additional purposes.
Total intensity source fluxes could better measure the continuum spectra of clumps  
in conjunction with independent data sets, improving the separation of different emission
 components and tightening constraints on dust emissivity and gas temperatures.
The maps provide upper limits to source polarization, allowing a statistical 
study of polarized contribution to anomalous emission similar to~\cite{dickinson2007}.
Finally, the catalog provides a cross-check of astrometry and absolute calibration
for instruments with access to the southern hemisphere.

\acknowledgements

This paper is dedicated to the memory of Andrew Lange, who gave wisdom 
and guidance to so many members of the astrophysics and cosmology 
community. His presence is sorely missed.
We thank our colleagues on the BICEP experiment for useful discussions.
QUaD is funded by the National Science Foundation in the USA, through
grants ANT-0338138, ANT-0338335 \& ANT-0338238, by the Science and
Technology Facilities Council (STFC) in the UK and by the Science
Foundation Ireland.  
We would like to thank the staff of the
Amundsen-Scott South Pole Station and all involved in the United
States Antarctic Program for the superb support operation which makes
the science presented here possible. Special thanks go to our intrepid
winter over scientist Robert Schwarz who spent three consecutive
winter seasons tending the QUaD experiment. The BOOMERanG
collaboration kindly allowed the use of their CMB maps for our
calibration purposes.  MLB acknowledges the award of a PPARC
Fellowship. SEC acknowledges support from a Stanford Terman
Fellowship. JRH acknowledges the support of an NSF Graduate Research
Fellowship and a Stanford Graduate Fellowship. CP and JEC acknowledge
partial support from the Kavli Institute for Cosmological Physics
through the grant NSF PHY-0114422.  EYW acknowledges receipt of an
NDSEG fellowship.
JMK acknowledges support from a John B.\ and Nelly L.\ Kilroy Foundation 
Fellowship.

\bibliographystyle{apj}
\bibliography{ms}

\begin{appendix}

\section{Simulations}
\label{app:sims}

The source extraction algorithm presented in Section~\ref{sec:srcextraction} 
is tested using simulated distributions of point sources, with and without a 
toy model diffuse background.
These simulations are used to determine the survey completeness, purity, the 
accuracy of recovered individual source parameters, the accuracy of recovered 
source distribution parameters, and the effect of choice of background smoothing
 kernel width $\sigma_{bck}$.

\subsection{Galactic Model Generation}
\label{subsec:simmodels}

The methods described in Appendices~\ref{subsubsec:simsrcdist} and~\ref{subsubsec:simdiff}
below are used to generate four types of galaxy simulation: random and correlated 
spatial distributions of point sources, with and without a diffuse background
component --- Table~\ref{tab:simtypesall} summarizes the properties of each simulation.

\subsubsection{Simulated Source Populations}
\label{subsubsec:simsrcdist}

Point sources are placed over the area of pixels occupied by the QUaD survey, using two 
methods to generate their spatial distribution.
The first is a simple random distribution, denoted Sim1, with a partner simulation 
(Sim2) also constructed from the same spatial distribution of sources but with the 
inclusion of a diffuse background component (see Appendix~\ref{subsubsec:simdiff} below).
The second more closely matches the observed clustering of discrete sources: 
A two-point angular correlation function $w\left(\theta\right)=k\theta^{-\gamma_{c}}$
is used to model the clustering of sources, and a power-law probability distribution 
function in galactic latitude $b$ is simultaneously implemented to capture the 
observation that sources tend to be concentrated towards the galactic plane, 
i.e. $p(b)\mathrm{d}b\propto b^{-\beta}\mathrm{d}b$. 
This simulation is called Sim3, and a further simulation, Sim4, is generated 
by taking the same point source population and adding a diffuse component.
The power-law exponents $\gamma_{c}$ and $\beta$ are chosen such that the spatial
distribution of sources qualitatively matches that in the QUaD data.
Since it contains both a diffuse background and spatially clustered point sources,
Sim4 is the model which most closely resembles the real data.

The physical properties of the sources are defined as follows: \flow\ total 
intensity source fluxes are drawn from a power-law model for the source counts 
($dN/dS\propto S^{\gamma_{S}}$, with $\gamma_{S}=-1.5$) between 0.1 and 250 Jy; the 
normalization is chosen to match the average source density (i.e. number per 
square degree) in the QUaD \I\ catalog.
Spectral indices between $100$ and \fhigh\ are generated using a gaussian probability
distribution function (p.d.f) of zero mean and unit width; \fhigh\ fluxes are 
generated by combining the spectral index and \flow\ flux.
The polarization fraction for each source is a random number drawn uniformly between 
0 and 20\%, while the polarization angle is also a uniform random number
 between 0 and $180^{\circ}$.

\subsubsection{Simulated Diffuse Background}
\label{subsubsec:simdiff}

Adding a model diffuse background allows its effect on recovered source properties 
to be assessed.
This component is modelled using a weighted sum of a single point source map 
smoothed to different resolutions.

Point source locations are generated using the correlation function approach described
in Appendix~\ref{subsubsec:simsrcdist} for Sim3 and Sim4, with a source density 
$\sim20$ times higher than the real data. 
All sources have the same \flow\ fluxes $S_{bck}$, a spectral index of 2.18 (the mean
spectral index of the diffuse emission as calculated in the Map Paper), and a
polarization fraction of 2\% in pure $+Q$ in galactic coordinates; these sources 
are placed within the boundaries of the survey in \I, \Q\ and \U\ maps at each 
frequency $\nu$, giving a set of template maps $m_{t,\nu,i}$, where $i$ refers to
each Stokes parameter.

To construct the model of diffuse emission, each $m_{t,\nu,i}$ is first smoothed with a 
set of circular gaussian kernels of differing width $\sigma_{j}$ to yield smoothed
 maps $m_{t,\nu,i,j}$, with $-1.4\leq\mathrm{log_{10}}(\sigma_{j})\leq0.2$ in intervals of
 $\delta\mathrm{log_{10}}(\sigma_{j})=0.2$.
The smoothed maps are then coadded with a different weight $w_{j}$ for each 
smoothing kernel, with $w_{j}\propto j^{7/4}$, yielding the diffuse model map $m_{bck,\nu,i}$:

\begin{equation}
m_{bck,\nu,i}=\sum_{j} w_{j} m_{t,\nu,i,j}.
\label{eq:diffsimsum}
\end{equation}
The $w_{j}$ amplitudes and $S_{bck}$ were chosen by requiring that the flux integrated over the whole
simulated map matched that in the data within $\sim10\%$.
A further consideration is that the relative weights between the $w_{j}$, and the exponent
 (defined somewhat arbitrarily here to be $7/4$), should be chosen such that substructure on 
different angular scales in the model matches the data; a power spectrum analysis may 
represent the best way of determining the $w_{j}$ but is beyond the scope of this paper.
Here, we merely note that with the choice of $\sigma_{j}$ and $w_{j}$ used above, the 
diffuse component of the simulations bear a qualitative resemblance 
to the data, as may be seen in Figure~\ref{fig:galsims}.

\subsubsection{Simulated Maps}
\label{subsubsec:simmaps}

\begin{figure*}[ht]
\resizebox{\textwidth}{!}{\includegraphics{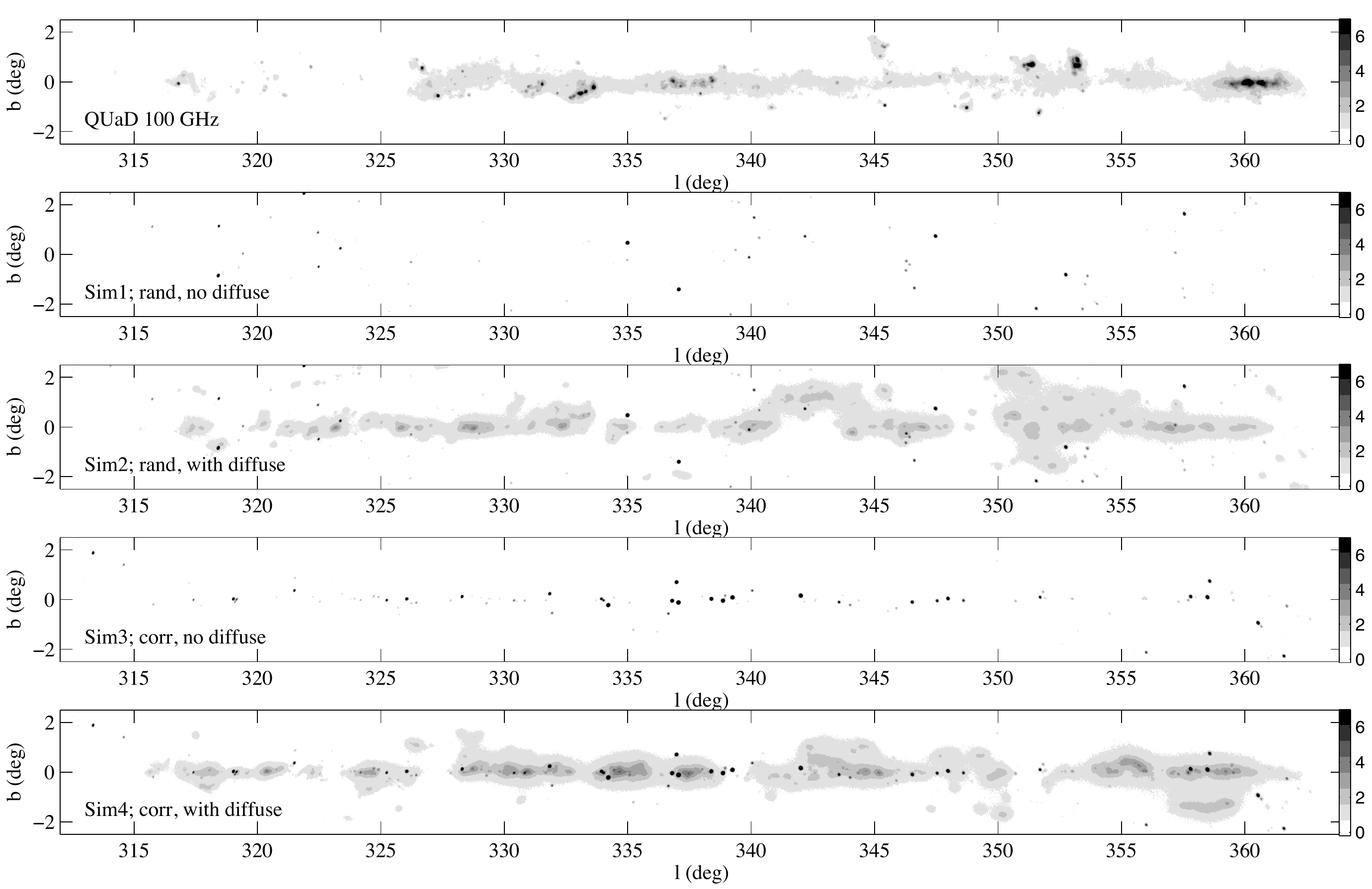}}
\caption{Comparison of QUaD \flow\ map to one realization of the simulations described in 
Appendix~\ref{subsubsec:simmaps}, with all maps transformed to galactic coordinates.
The color scale is MJy/sr.
From top to bottom: QUaD data, random point sources only, random sources plus
diffuse background, spatially correlated sources, and spatially correlated 
sources plus diffuse background.
}
\label{fig:galsims}
\end{figure*}

\begin{deluxetable}{c|cc|cccc|cccc}
\tabletypesize{\scriptsize}
\setlength{\tabcolsep}{0.06in}
\tablecolumns{9}
\tablewidth{0pt}
\tablecaption{Simulation Summary \label{tab:simtypesall}}
\tablehead{
\colhead{Simulation}  &
\colhead{Source spatial}     &
\colhead{Diffuse } &
\multicolumn{4}{c}{90\% Completeness \hspace{2cm}} &
\multicolumn{4}{c}{Purity\tablenotemark{a} \hspace{2cm}} \\
\colhead{} & 
\colhead{distribution} & 
\colhead{emission} & 
\colhead{$I_{100}$ (Jy)} & 
\colhead{$I_{150}$ (Jy)} & 
\colhead{$P_{100}$ (Jy)} & 
\colhead{$P_{150}$ (Jy)} &
\colhead{$p_{I,100}$} & 
\colhead{$p_{I,150}$} & 
\colhead{$p_{P,100}$} & 
\colhead{$p_{P,150}$} 
}
\startdata
Sim1 & random     & no  & 1.6 & 1.5 & 1.2  & 0.9 & 0.88 & 0.89 & 0.90 & 0.91 \\     
Sim2 & random     & yes & 1.5 & 1.5 & 1.3  & 0.9 & 0.43 & 0.40 & 0.93 & 0.91 \\ 
Sim3 & correlated & no  & 5.1 & 2.4 & 13.7 & 1.2 & 0.98 & 0.98 & 0.97 & 0.92 \\
Sim4 & correlated & yes & 5.9 & 2.9 & 20.3 & 1.1 & 0.46 & 0.43 & 0.93 & 0.94 \\
\enddata 
\tablenotetext{a}{Purity estimated at point source signal-to-noise threshold of 5 (3)
in total (polarized) intensity.}
\end{deluxetable}

Having determined the positions and physical properties of the sources, they are then
placed in a `source map', $m_{src}$, of the same size and pixel resolution as the QUaD maps.
A similar map $m_{bck}$ is generated for the diffuse background following 
Appendix~\ref{subsubsec:simdiff}; $m_{src}$ alone is the input sky for Sim1 and Sim3, with
 $m_{src} + m_{bck}$ used for Sim2 and Sim4.
For each type of point source spatial distribution (random or correlated), the 
coordinates of each source are the same with and without a diffuse component present.
This allows the effect of the background on source fluxes to be investigated separately
from spatially correlated source positions.
Note that a different realization of the diffuse component is used for the random and
correlated source simulations, as may be seen in Figure~\ref{fig:galsims}.

Simulated detector timestream is interpolated from these maps using the pointing 
information from each day of real QUaD data, and realistic noise added as described 
in the Map Paper.
The simulated signal+noise data is then subjected to the same filtering and 
mapmaking steps as in the QUaD data pipeline described in the Map Paper.
Figure~\ref{fig:galsims} shows the QUaD \flow\ \I\ map, and maps from one realization 
of each simulation type.

The maps are passed through the source extraction algorithm described in 
Section~\ref{sec:srcextraction} to generate catalogs of sources in \I, \Q\ and 
\U\ at the two QUaD frequencies.
To determine quantities such as spectral index and polarization fraction (that is, 
quantities which require the catalog from more than one map), sources are matched 
using the simple spatial criterion described in Section~\ref{sec:srcextraction}.
Many sky realizations are processed to build up sufficient statistics to 
characterize the survey depth and systematic effects.

\subsection{Completeness}
\label{subsec:completeness}

\begin{figure}[h]
\resizebox{\columnwidth}{!}{\includegraphics{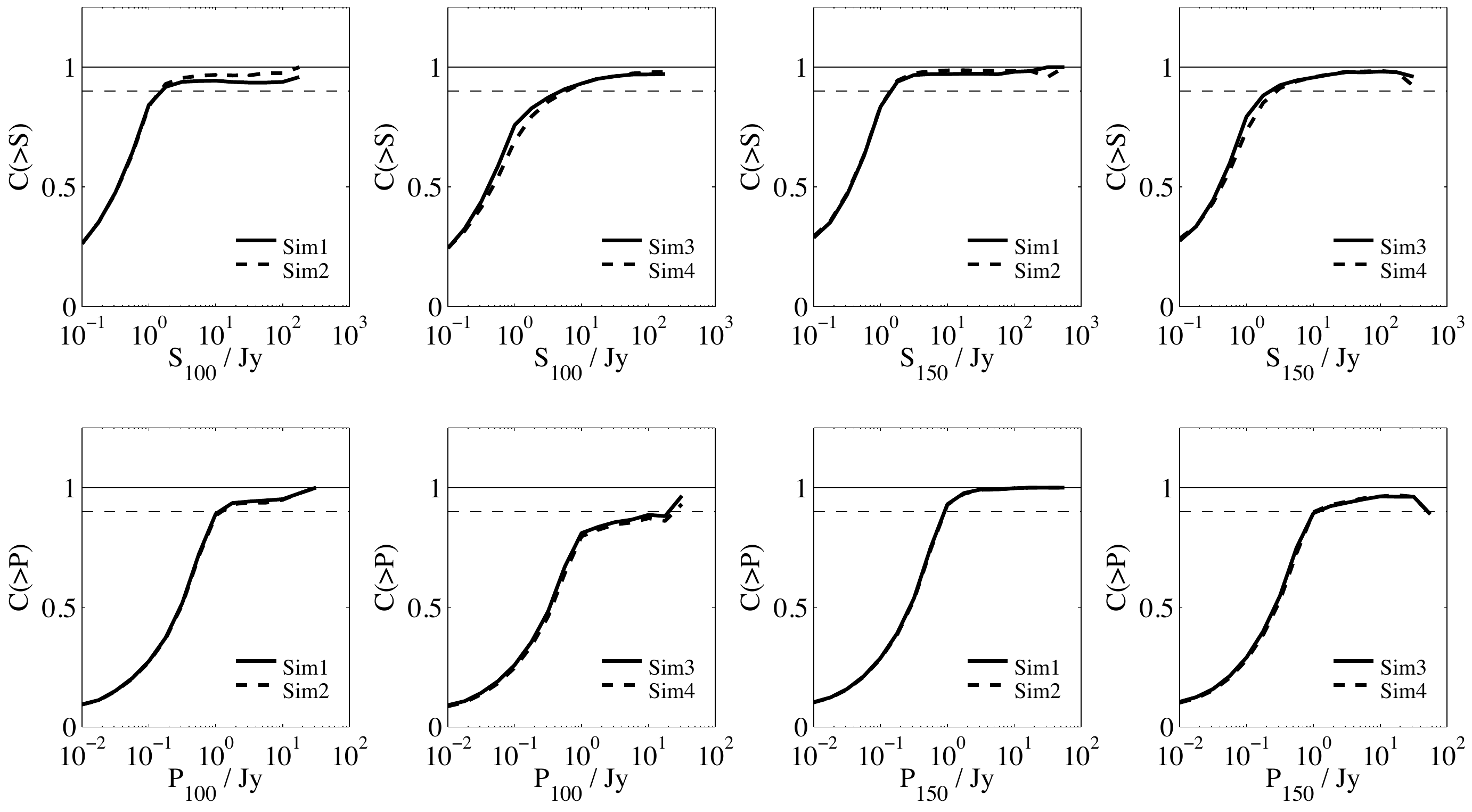}}
\caption{Completeness estimated from simulated point source distibutions in total (top)
and polarized (bottom) intensity.
The left pair of columns is for \flow, right pair for \fhigh.
Within each pair, the left column is for randomly distributed sources with and
without a diffuse component, and likewise for spatially correlated sources in 
the right column.
Solid black curves correspond to sources with no diffuse background 
present, dashed black curves are for the same source distributions with 
diffuse emission added.
The thin dashed black line indicates 90\% completeness.
}
\label{fig:completenesssims}
\end{figure}

The completeness $C(>S)$, the fraction of input sources recovered above flux $S$, 
is shown in Figure~\ref{fig:completenesssims} for the QUaD frequency bands 
for each set of simulations. 
The 90\% completeness limits (the flux $S_{90}$ at which the completeness
reaches 90\%) for the simulated catalogs are summarized in 
Table~\ref{tab:simtypesall}.

The 90\% completeness limit in total intensity is generally higher than in 
polarized intensity; this is in part due to the higher signal-to-noise detection 
threshold than that used in polarization (5 compared to 3).
The limit in \I\ is also increased due to the shape of the completeness
curve, where $C(>S)$ increases rapidly to 0.5 in \I\ but then approaches unity 
more slowly --- see Figure~\ref{fig:completenesssims}.
This effect is attributed to residual $1/f$ noise in the \I\ map, rather than the diffuse
background or source confusion, since it is present for all simulations.
In Sim1 and Sim2, the 90\% completeness limits for \I\ are 1.6 and 1.5 Jy respectively
at \flow, and 1.5 Jy for both simulation types at \fhigh.
The 90\% limit is higher for Sim3, reaching 5.1 (2.4) Jy at 100 (150) GHz.
Since this simulation contains no background, the effect is solely due to increased 
source confusion resulting from the spatial clustering of the sources.
For Sim4, these numbers increase marginally to 5.9 and 2.9 at 100 and \fhigh\ 
respectively, indicating that the spatial clustering of sources is more important
than the presence of a diffuse background.

In polarization, the diffuse background is faint enough that it does not strongly
affect the completeness as shown in Figure~\ref{fig:completenesssims}.
A more important effect is the spatial clustering of sources --- at \flow, the survey
is only 90\% complete in polarization by 13.7 Jy for Sim3 and 20.3 Jy for Sim4.
At \fhigh, these figures drop to 1.2 and 1.1 Jy respectively.
This is likely due to three effects.
First, the survey coverage is smaller at \flow\ due to the smaller area of
the QUaD focal plane at this frequency (the \flow\ survey is approximately 7\%
smaller in polarization than \fhigh).
As a result, some input sources may not lie in the \flow\ survey area; however,
for spatially correlated sources, which are clustered close to the plane of the
galaxy, this is likely a small effect.
Second, the larger beam at \flow\ causes source confusion.
The randomly oriented polarization angles could result in polarized flux
dilution, reducing the completeness.
Figure~\ref{fig:completenesssims} shows that this effect is reduced at
\fhigh, in support of this idea.
Third, the completeness is a function of polarization fraction and total flux.
No source with zero polarization fraction, or no faint source with low polarization
fraction, will therefore be detected.
Since the input source polarization fractions lie between 0 and 20\%, this effect
could also contributed, though it would be expected at both frequencies.
In summary, in all simulations except \flow\ Sims 3 and 4, the 90\% point source completeness 
limit in polarization is $0.9$--$1.3$ Jy.

\subsection{Purity}
\label{subsec:purity}

\begin{figure}[h]
\resizebox{\columnwidth}{!}{\includegraphics{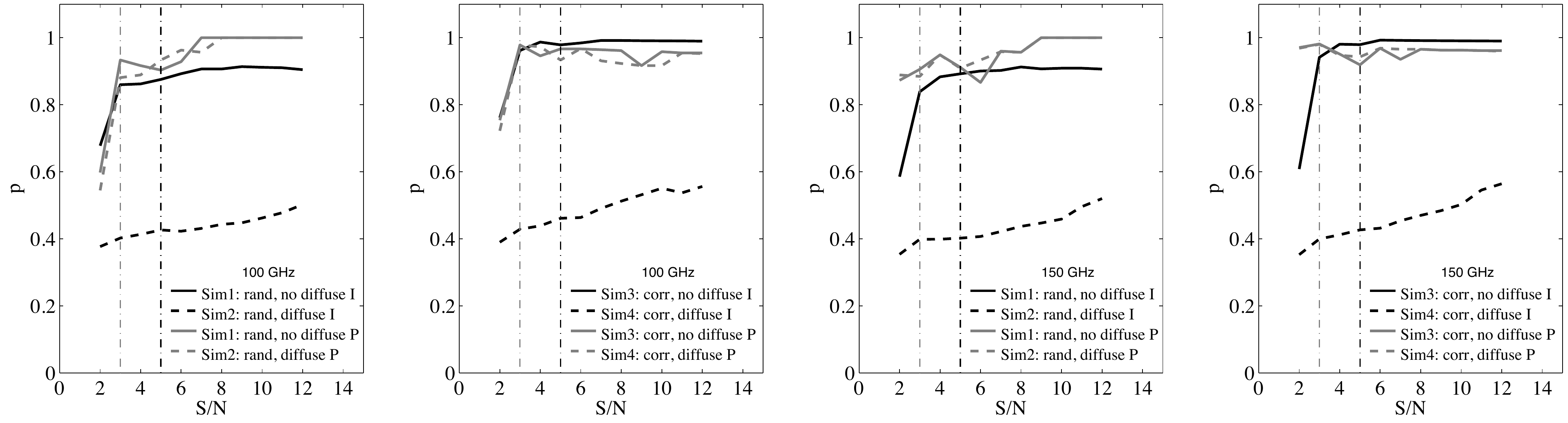}}
\caption{Purity as estimated from simulated point source distributions as a function
of signal-to-noise (S/N).
The left (right) pair of columns correspond to purity at 100 (150) GHz.
Within each pair, the left plot is for randomly distributed sources with and
without a diffuse background (Sim1 and Sim2), and the right plot similar but for spatially 
correlated sources (Sim3 and Sim4).
Survey purity in total (polarized) intensity are plotted in black (gray). 
The vertical black dot-dashed lines correspond to the S/N ratio used to extract sources 
in \I; the vertical gray dot-dashed line is the same for polarization.}
\label{fig:purity}
\end{figure}

The `purity' of the survey $p$ is the number of recovered sources which were matched 
spatially to the input catalog divided by the total number of recovered sources.
This is quantified by comparing the input and recovered source catalogs in Sims1--4.
Figure~\ref{fig:purity} shows the purity as a function of signal-to-noise threshold 
in both total and polarized intensity.
The values of $p$ at the chosen extraction thresholds ($\mathrm{S/N}>5$ in \I, 
$\mathrm{S/N}>3$ in $P$) are summarized in Table~\ref{tab:simtypesall}.

\begin{figure}[h]
\resizebox{\columnwidth}{!}{\includegraphics{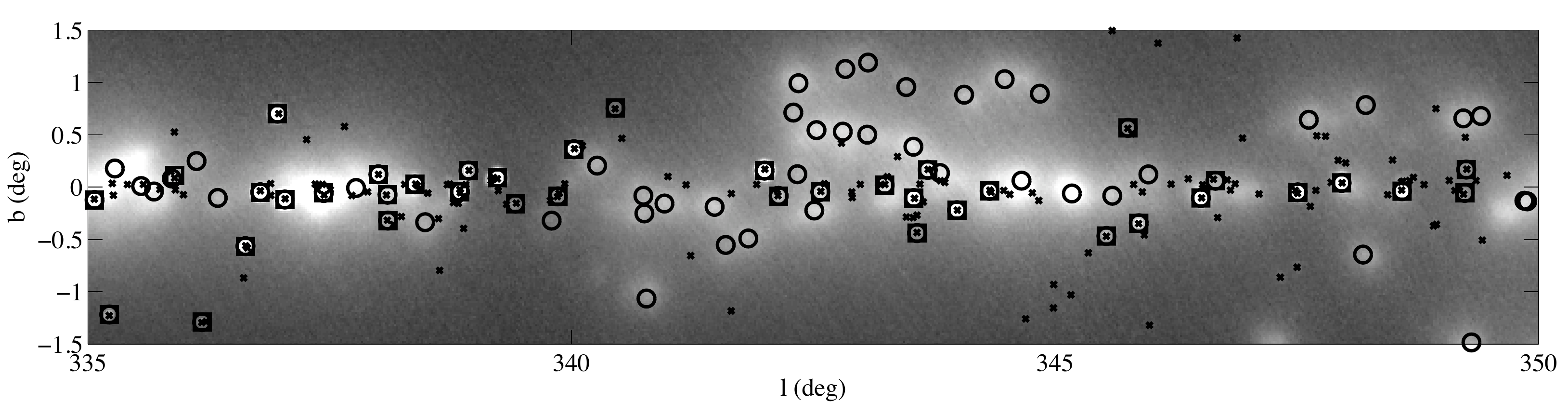}}
\caption{Illustration of purity in the presence of diffuse background.
The image is a subsection of the Sim4 \flow\ \I\ map shown in the bottom panel of
Figure~\ref{fig:purity}. 
Crosses correspond to the locations of input sources, squares are the sources detected from 
Sim3, and circles are the sources detected in Sim4 (recall that the input point source populations 
are identical for Sim3 and Sim4).
In Sim4, false sources are detected from substructure in the diffuse emission --- these sources
have a circle that does not enclose a cross.
This systematic effect is the cause of the low purity for simulations which have a diffuse
background present, as described in Appendix~\ref{subsec:purity}.}
\label{fig:purity_diff}
\end{figure}

Between detection thresholds of $2<\mathrm{S/N}<8$, the purity increases more rapidly with 
$\mathrm{S/N}$ in both total and polarized intensity if the spatial distribution of sources is 
correlated rather than random.
This is due to flux boosting of a source by fainter, spatially coincident sources which are not 
resolved themselves.
In total intensity, the purity appears poor (40-50\%) when the diffuse background is present (Sim2 and Sim4),
 even at a detection threshold of $\mathrm{S/N}>12$.
However, these `false detections' are not noise fluctuations: Investigation of the output catalogs from 
Sim2 and Sim4 demonstrated that the detected sources unmatched to the input catalog were 
beam-scale or extended sources associated with substructure in the simulated diffuse background.
Figure~\ref{fig:purity_diff} illustrates this effect for a Sim4 realization, demonstrating how substructure
can be erroneously detected as real discrete sources.
This effect makes `purity' an ambiguous concept in the context of separating sources from a diffuse background
with enough power on beam-sized angular scales:
If the signal-to-noise of the diffuse background is comparable to that of the discrete sources, the
purity is not dominated by noise fluctuations, but the inability to distinguish the sources
of interest from resolved substructure in the diffuse emission. 
Simulations Sim2 and Sim4 fall into this category, which is why their purity is low even at high signal-to-noise
thresholds; this effect is simply a consequence of the model parameters chosen in Appendix~\ref{subsubsec:simdiff}.
As shown in Section~\ref{sec:data}, 97\% (87\%) of the QUaD sources detected in total intensity
at 100 (150) GHz have IRAS-PSC counterparts; this indicates that the QUaD catalog is likely very 
pure, and that more beam-scale substructure is present in Sim2 and Sim4 than in the real data.
A power-spectrum approach, advocated in Appendix~\ref{subsubsec:simdiff}, represents the best way
to determine the amount power on different angular scales.

The diffuse polarized background is faint enough that there is little difference between the 
purity in each type of simulation.
Figure~\ref{fig:purity} shows that a polarized source detection threshold of $\mathrm{S/N}>3$ 
results in a catalog that is $\sim90\%$ pure for randomly distributed sources, or $\sim100\%$ pure 
for spatially correlated sources.

\subsection{Source Recovery}
\label{app:srcdist}

Recovered catalogs from the four sets of simulations are used to estimate 
how accurately the input parameters of individual sources and source distributions
 can be recovered.
While the output distributions for Sim1 and Sim3 (those without a diffuse background)
should be insensitive to the choice of input distribution parameters, such
as the source counts slope $\gamma_{S}$, the systematic biases introduced by the 
background (simulations Sim2 and Sim4) do depend on the background model parameters 
$\gamma_{c}$, $\beta$, $w_{i}$, $\sigma_{i}$, and the overall amplitude.
We therefore caution that while the toy model of the diffuse component allows
a qualitative impression of how source properties can be corrupted, the amount of
corruption depends on these parameters to an extent that may differ from the real
data.

\subsubsection{Recovery of Individual Source Parameters}
\label{app:srcrecovery}

\begin{figure*}[ht]
\resizebox{\textwidth}{!}{\includegraphics{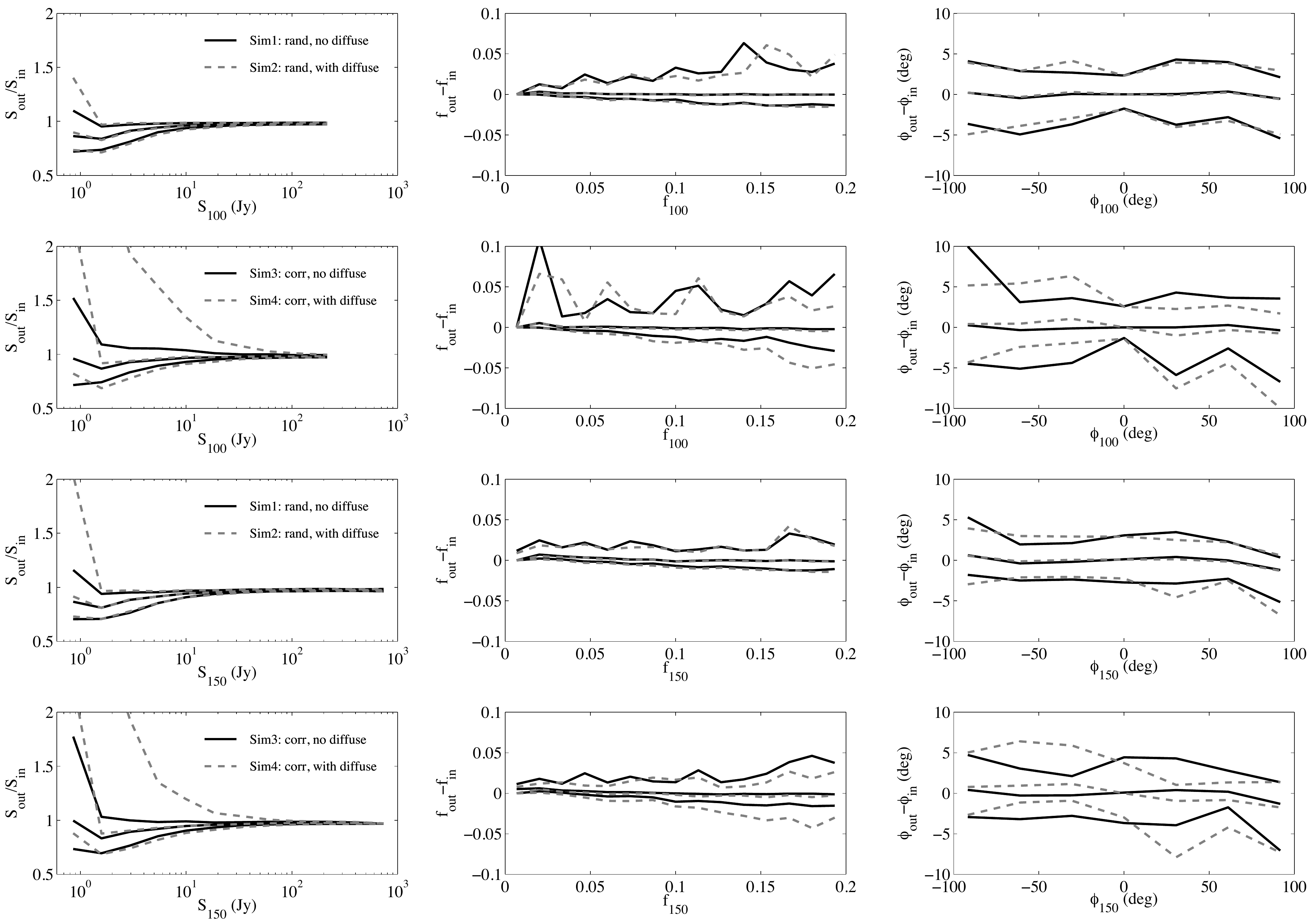}}
\caption{Recovery of simulated point source properties as a function of the input. Left
to right, the columns are total intensity source flux $S$, polarization fraction $f$,
and polarization angle $\phi$.
$S$ is expressed as a ratio between the output and input values, while $f$ and $\phi$ 
are expressed as the difference of the two quantities.
Top two rows are \flow, while lower two rows are \fhigh. 
Within each pair of rows at a fixed frequency, the simulations used are Sim1 and Sim2 in the first
row, and Sim3 and Sim4 in the second row.
The contours indicate the 16, 50 and 84th percentiles of each quantity as calculated
from simulations.
Solid black contours are for simulations without a diffuse background (Sim1 and Sim3), while gray dashed 
contours indicate the simulations with diffuse emission (Sim2 and Sim4).
}
\label{fig:simrecov}
\end{figure*}

Figure~\ref{fig:simrecov} compares the recovered source properties to their input values.
In total flux $S$, we show the 16, 50 and 84\% percentiles of the ratio $S_{out}/S_{in}$, while
the same percentiles are shown for polarization fraction difference $f_{out}-f_{in}$ and 
polarization angle difference $\phi_{out}-\phi_{in}$.
Both \flow\ and \fhigh\ simulations are shown, though similar behaviour is observed at both frequencies.

For total intensity $S$, in all simulations the median $S_{out}/S_{in}$ falls as sources get 
fainter, but is within a few percent of unity down to 10 Jy (comparable to the absolute calibration 
uncertainty of 3.5\%).
At this flux, the 16 and 84\% percentiles of $S_{out}/S_{in}$ are $\sim5\%$ from the median for
simulations without a diffuse background (Sim1 and Sim3).
Below 10 Jy, rather than more numerous faint sources being boosted to higher fluxes, the systematic
defecit in recovered flux is due to filtering of the timestream before map coaddition (see Map
Paper); this demonstrates that filtering effects are more important than flux boosting due to 
instrumental noise at low fluxes.

Including diffuse emission results in a wider and more asymmetric distribution of recovered 
flux due to background contamination; at 10 Jy in Sim3 (no diffuse emission), the systematic uncertainties are $\sigma_{-}=0.04$
and $\sigma_{+}=0.07$, while the same quantities in Sim4 (including diffuse emission) are
$\sigma_{-}=0.07$ and $\sigma_{+}=0.36$, where $\sigma_{-}$ is the difference between the 50th and
16th percentiles of the ratio $S_{out}/S_{in}$ at 10 Jy, and $\sigma_{+}$ is the difference 
between the 84th and 50th percentiles.
The asymmetric errors due to background contamination are discussed further in~\ref{app:counts}
in the context of the source counts, $dN/dS$.

The recovery of point source polarization fraction is not strongly biased in any of the simulations.
The median difference between input and output polarization fractions is $<1\%$, at which point 
beam systematic effects become important (see Instrument Paper for details).
The scatter on $f_{out}-f_{in}$ generally increases with higher polarization fraction, rising 
from $\sim1\%$ at $f\sim1\%$ to 5\% for $f=20\%$. 
This is because only the brightest sources tend to be detected at low $f$, while at high $f$ 
both bright and faint sources are included, increasing the variance of $f_{out}-f_{in}$.
The distribution of $f_{out}-f_{in}$ is skewed towards positive values; this effect is attributed to 
the addition of noise to the total polarized flux as $\sigma^{2}_{P}=\sigma^{2}_{Q}+\sigma^{2}_{U}$.
No systematic bias is introduced when a diffuse component is present, indicating that the background
removal strategy is effective for determining fluxes of polarized sources.

One might ask why the recovered $f$ is not systematically lower than the input, since the polarization
angles assigned to simulated sources are random and therefore should average to zero when sources are 
confused, as in Sim3 and Sim4?
The reason is that the simulations use a power-law distribution of fluxes, resulting in many more 
faint sources per unit solid angle than bright sources.
In a given resolution element, faint sources will largely cancel each others' polarized flux, while
a statistically unlikely (but far brighter) polarized source will dominate the polarized emission.

The simulations show that polarization angle difference $\phi_{out}-\phi_{in}$ suffers a 
systematic shift of $<1^{\circ}$, with 16 and 84 percentiles less $<5^{\circ}$.
The scatter in $\phi$ is not strongly affected by a diffuse component, but increases marginally 
when sources are spatially correlated.
This increased scatter is due to confusion of sources with random polarization angles within
a single beam element; although on average fainter sources with random $\phi$ will average to
zero, their presence will introduce extra fluctuation into the polarization angle of the
brightest source.

\subsubsection{Recovery of Source Counts}
\label{app:counts}

\begin{figure}[h]
\resizebox{\columnwidth}{!}{\includegraphics{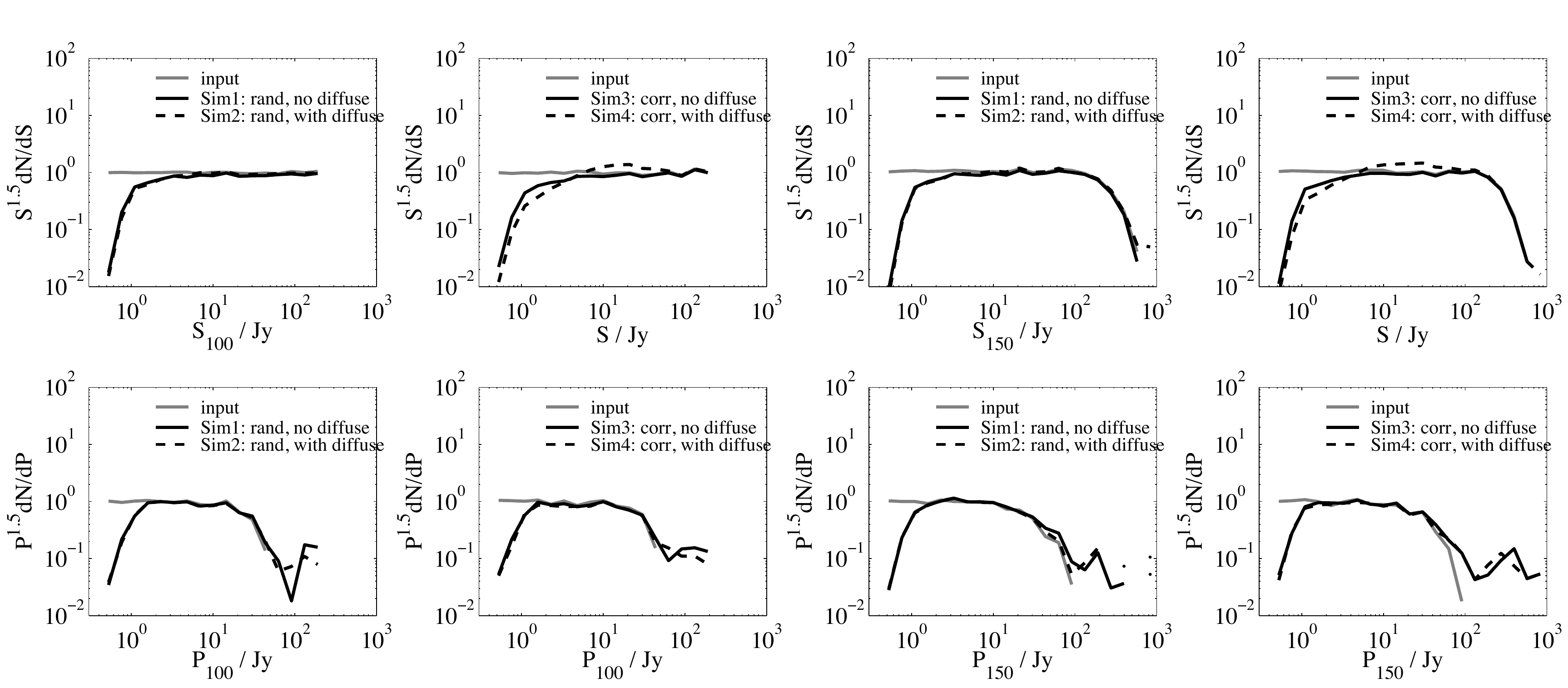}}
\caption{Source counts $S^{1.5}dN/dS$ from simulated point source distibutions.
Input source counts are gray, solid black is point-source only
simulations (Sim1 and Sim3), dashed black is simulations including a diffuse background (Sim2 and Sim4).
Top row is total intensity, bottom row is polarized intensity.
Left pair of columns is \flow, right pair is \fhigh.
Within each pair of columns, the left column is randomly distributed sources (Sim1 and Sim2), 
right is correlated source locations (Sim3 and Sim4).
The steepening of the \fhigh\ counts at high flux is due to the convolution of
a power-law flux distribution at \flow\ with a gaussian spectral index distribution.}
\label{fig:dNdSsims}
\end{figure}

Of particular interest to source surveys are the source counts $dN/dS$; it is 
therefore important to address whether this quantity can be accurately recovered
in the presence of noise and a diffuse background.
Figure~\ref{fig:dNdSsims} shows $S^{1.5}dN/dS$ as a function of source flux, for
the input and recovered source distributions, in each of the simulation types Sim1--Sim4.
Since the input counts were $dN/dS \propto S^{-1.5}$, in this plot perfectly
recovered counts appear as a line of zero gradient.
The Figure shows that for Sim1--Sim3, the counts obey the expected 
property of being well-recovered at high flux, but falling off as the flux
approaches the survey detection threshold.
Note that the fall-off at high flux at \fhigh\ is due to the finite maximum \flow\ 
source flux convolved with the assumed spectral index distribution; this part of 
the input counts does not obey a power-law but is well recovered by the source extraction.
The source counts show similar properties at both frequencies, in total and polarized
intensity.

Only in the case of the total intensity fluxes of clustered sources in the presence
of a diffuse background (Sim4) do we see significant deviations from the ideal behavior; 
in this case, boosting occurs from low fluxes to higher fluxes.
This may be seen from the defecit of sources at low flux relative to simulations without
a diffuse component, and an excess above the input counts at high fluxes.
In the high flux regime, where source counts are typically fit, the shape of 
$S^{1.5}dN/dS$ is heavily distorted from a simple power-law, rendering constraints
on this quantity difficult to measure.
Note the distinction between this type of flux boosting, which is due to background 
emission, and that due to uniform survey noise; in extragalactic surveys of radio 
sources, only the latter is normally considered~\citep[e.g.][]{muchovej2010}, and 
can be corrected by marginalizing over the underlying source count parameters.
For the former, which is termed `background boosting' here, de-boosting source 
fluxes is not an easy problem.
While diffuse emission such as the Cosmic Microwave Background (CMB) is very well
 characterized as a gaussian random noise distribution (and could potentially also
 be marginalized over to find the true source flux), the properties of the diffuse
 galactic emission (such as morphology, spectral behavior, and projection effects 
along different lines of sight through the galaxy) are at present poorly constrained
 and not easy to model.
Contamination by galactic emission is also asymmetric in the sense that
sources are only ever boosted to higher fluxes, because unlike CMB fluctuations 
(unpolarized) galactic emission is always positive.

We therefore caution against over-interpretation of the source count slope in the 
presence of a galactic `background', since the slope is shown by simulations to be
corrupted despite the aggressive background filtering.
The level of corruption is dependent on the parameters of the diffuse background model 
(such as the amount of power in beam-scale substructure).
As we only use one set of parameters in the simulations, the results presented 
above are not intended to precisely quantify this systematic error, but to explore 
how the source counts can be affected by the background in a restrictive region of
diffuse model parameter space.
Despite these limitations, the source counts of the QUaD survey data shown in 
Figure~\ref{fig:dNdS} do appear to obey a power-law at high fluxes, indicating
that in this regime background contamination is likely unimportant.

\subsubsection{Recovery of Spectral Index Distribution}
\label{app:specinddist_recovery}

\begin{figure}[h]
\resizebox{\columnwidth}{!}{\includegraphics{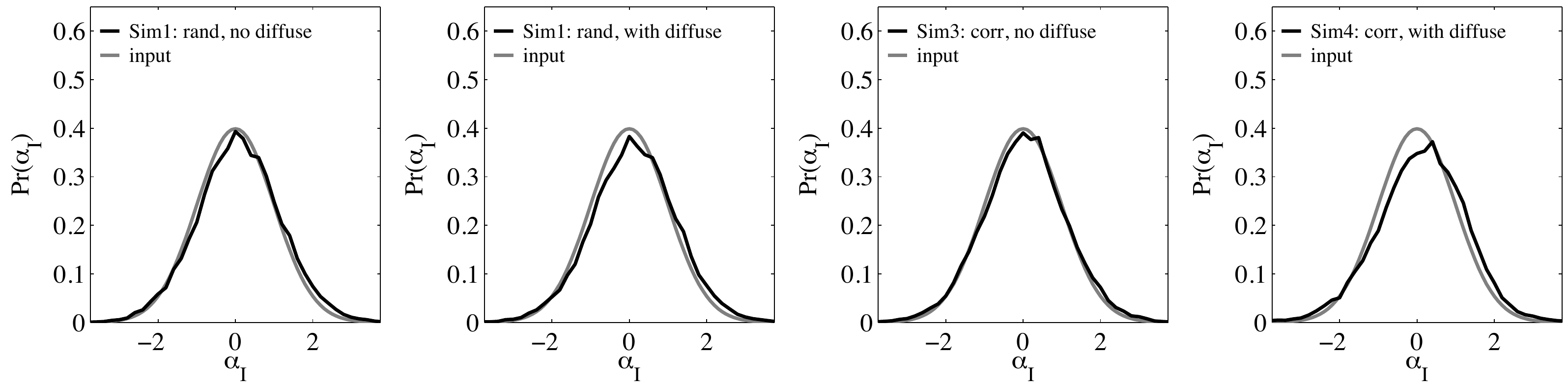}}
\caption{Recovered total intensity source spectral index distribution shown in gray for 
(left to right) Sim1, Sim2, Sim3 and Sim4, compared to input (black).}
\label{fig:specinddist_recovery}
\end{figure}

Figure~\ref{fig:specinddist_recovery} shows the recovered total intensity 
spectral index probability distribution $Pr(\alpha_{I})$ for one realization of each 
simulation type.
The input distribution, a gaussian p.d.f of unit rms, is well recovered for randomly 
distributed and spatially correlated sources without a diffuse background (Sims 1 and 3).
Sims 2 and 4 exhibit a small shift of the distribution center towards larger values; this is
likely due to contamination of faint source fluxes by the diffuse emission.
In support of this notion, the shift is larger for Sim4 than Sim2 since more sources lie
close to the plane of the galaxy, where diffuse emission is brightest.

We conclude that the recovered spectral index distribution from the QUaD data 
(Section~\ref{subsec:dataspecdist}) may be biased slightly high as a result of background
contamination preferentially affecting the \flow\ data.
However, the degree to which $Pr(\alpha)$ is corrupted in simulations depends
on the diffuse model input parameters; we therefore refrain from quantifying
the effect.

\subsubsection{Recovery of Correlation Function}
\label{app:corrfunc}

\begin{figure}[h]
\resizebox{\columnwidth}{!}{\includegraphics{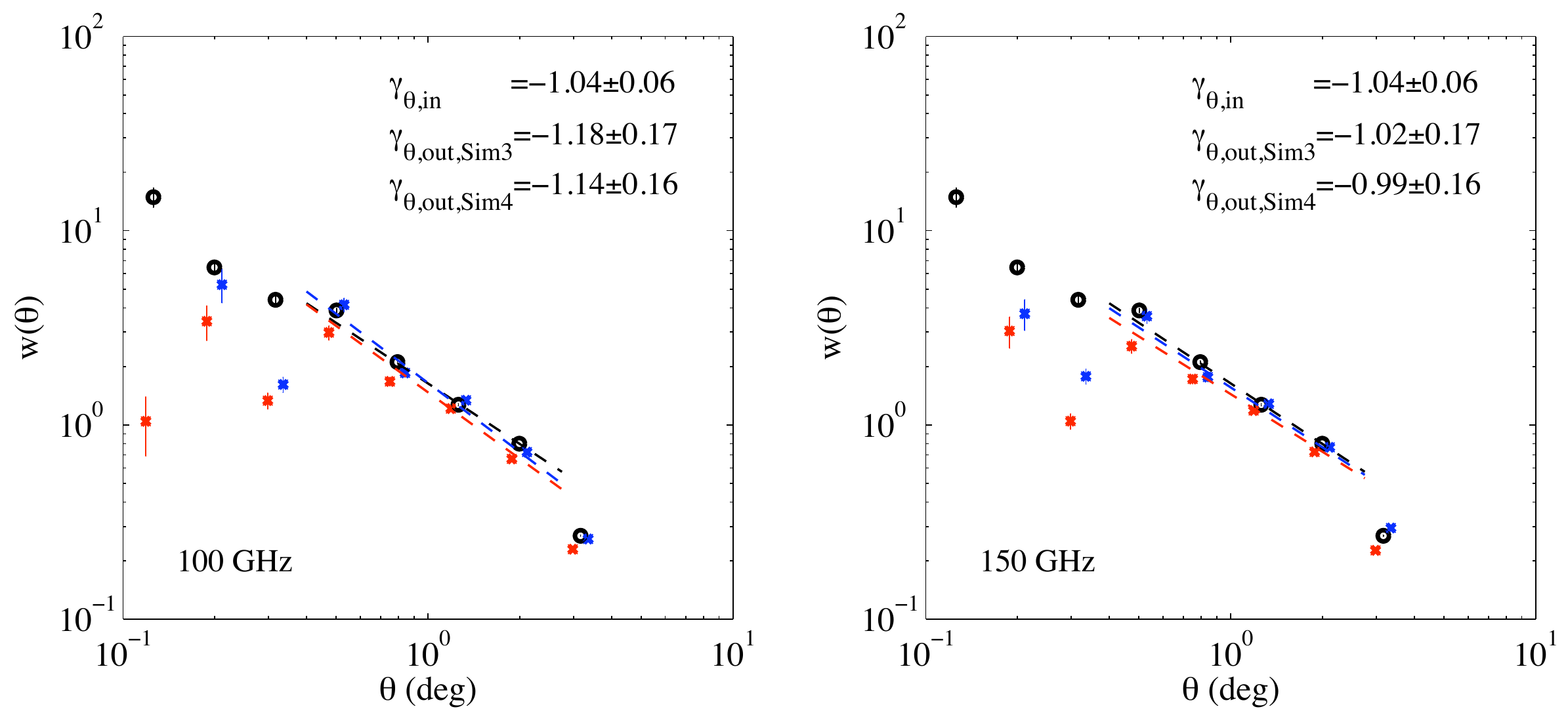}}
\caption{Recovery of input source correlation function $w(\theta)$ as a function of 
source separation $\theta$.
Input from one realization of the galaxy model is shown as black circles, with
the recovered function shown as crosses for one realization of Sim3 (blue) and Sim4
(red), with small horizontal offsets applied for clarity. 
Left is \flow, right is \fhigh. 
The deviation from a power-law at $\theta>2^{\circ}$ is due
 to the anisotropic source position distribution; this also occurs in the input
 simulations. 
At $\theta<0.4^{\circ}$, the power-law distribution is not well recovered 
--- see text for an explanation. 
The recovered values of the power-law slope between $0.4^{\circ}<\theta<2^{\circ}$ are
shown in the upper right of each panel.
In this range, where survey shape effects are insignificant,
the input slope is well-recovered.
}
\label{fig:simclustering}
\end{figure}

Simulations Sim3 and Sim4 are used to test the recovery of the input source
spatial distribution parameters for sources clustered in the plane of the
galaxy, as described in Section~\ref{subsec:sourceclustering}.
The angular correlation function is constructed as in Equation~\ref{eq:angcorr},
except this time $H_{d}(\theta)$ (the number of sources with a neighbor at 
separation $\theta$) is derived from a realization of Sim3 or Sim4, depending 
on whether a diffuse background is included.
As with the real data, Sim1 is used to generate a histogram of the number of 
randomly distributed sources with a neighbor at separation $\theta$, $H_{r}(\theta)$.
Correlation functions are generated for clustered point sources with and
without a diffuse background present; the results for a single simulated 
realization are presented in Figure~\ref{fig:simclustering}.

The black input points show that while a power-law correlation function is
traced at small $\theta$, at large source separations the slope becomes 
steeper.
This is a result of the anisotropic nature of the source distribution; less 
than 1\% of sources at each frequency lie further than $3^{\circ}$
from the plane, and those inside are preferentially located towards $b=0$ (see
Section~\ref{subsec:bdist}).
Only source clustering in galactic longitude contributes to the 
probability of finding a source separation greater than $\sim3^{\circ}$, 
suppressing the correlation function at large angular scales.
The reconstructed $w(\theta)$ from the simulated data shows this effect, 
demonstrating that although the correlation function is intrinsically 
suppressed due to the distribution of sources in the galaxy, it is 
still well-recovered at large angular separations.  
At angular scales $<0.4^{\circ}$, $w(\theta)$ becomes poorly recovered.
This is due to the large probability that the closest neighbors to bright, 
rare sources are faint and likely below the detection threshold.
Therefore $w(\theta)$ is only well-recovered when the source separation 
is large enough that the probability of a bright neighboring source is 
significant.

Figure~\ref{fig:simclustering} shows that the slope of the correlation function
is well-recovered in the range $0.4^{\circ}<\theta<2^{\circ}$ to within the
uncertainties; this is the range chosen for fitting $w(\theta)$ in the QUaD
data (Section~\ref{subsec:sourceclustering}).

\subsection{Effect of Background Kernel $\sigma_{bck}$}
\label{subsec:sigmabck}

Without removal of the diffuse background, the measured flux of each 
extracted source in the QUaD survey can be heavily influenced by proximity
to other bright sources and/or diffuse emission; either can add excess signal 
when source fluxes are determined, biasing recovered fluxes high.
The filtering scheme described in Section~\ref{sec:srcextraction} is designed to 
suppress the background by subtracting a template of diffuse emission from the maps.
The template is a smoothed version of the raw survey map, with point source 
pixels replaced by their local median.
Constructing the template requires a choice of smoothing scale $\sigma_{bck}$, which
represents the minimum angular scale on which background fluctuations are
assumed significant.
The results from the QUaD survey and the simulations in 
Appendices~\ref{subsec:completeness}--\ref{app:srcdist} are dependent on the choice of
$\sigma_{bck}$; here we investigate the effect of varying the value of this parameter.

\begin{figure}[h]
\resizebox{\columnwidth}{!}{\includegraphics{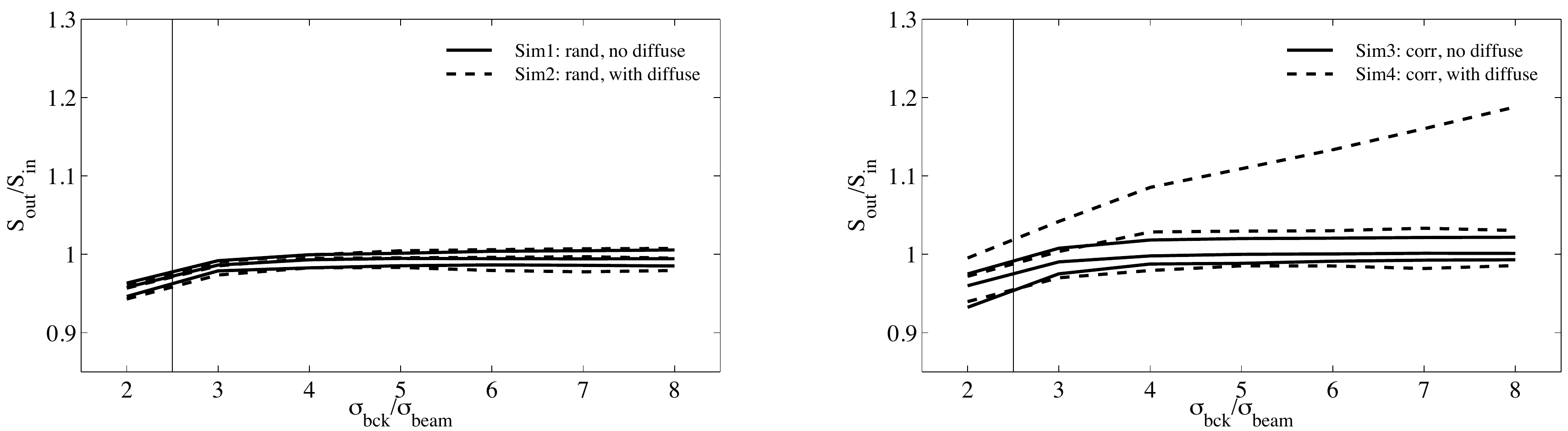}}
\caption{Effect of $\sigma_{bck}$ on recovered total intensity
flux $S$ for simulated point sources above 25 Jy at \flow.
Solid lines correspond to simulations without a diffuse background, while dashed
lines include a model of diffuse emission.
Left: Simulations of randomly distributed point sources.
Right: Simulations of clustered point sources.
For each simulation, the 16th, 50th and 84th percentiles of $S_{out}/S_{in}$ are plotted
as a function of $\sigma_{bck}/\sigma_{beam}$.
The solid vertical line corresponds to the value used on the QUaD data, $\sigma_{bck}/\sigma_{beam}=2.5$,
a compromise between excess filtering of source flux (low $\sigma_{bck}$) and
excess background contamination (large $\sigma_{bck}$).}
\label{fig:kernwidth}
\end{figure}

Figure~\ref{fig:kernwidth} shows the ratio of recovered \flow\ \I\ fluxes to the input
 as a function of $\sigma_{bck}/\sigma_{beam}$ for sources above 25 Jy (similar 
results are found at \fhigh).
The left panel shows that for Sim1, source fluxes are recovered to within $\sim2\%$ 
if $\sigma_{bck}/\sigma_{beam}>3$, with a defecit of $5\%$ by $\sigma_{bck}/\sigma_{beam}=2$.
A similar result is found for Sim2, though increased variance is found for
$\sigma_{bck}/\sigma_{beam}>3$.
The systematic reduction of flux at low $\sigma_{bck}/\sigma_{beam}$ is due to source
flux being subtracted with the background as $\sigma_{bck}\rightarrow\sigma_{beam}$.
Increased variance is found when source locations are correlated, as in Sim3 (see 
right panel of Figure~\ref{fig:kernwidth}).
Since this simulation is devoid of background emission, the larger 84th percentile is due
 to source confusion, as the likelihood of more than one source per beam is larger
for spatially correlated sources. 
As might be expected, the 16th percentile is not significantly changed compared to 
randomly distributed sources since confusion cannot reduce recovered source fluxes.
We note that despite source confusion, fluxes are still generally recovered to within 
$5\%$ or better, independent of $\sigma_{bck}$.
Including a diffuse background (Sim4) results in variance of $S_{out}/S_{in}$ 
which increases as a function of $\sigma_{bck}/\sigma_{beam}$.
The variance is skewed towards positive fluctuations due to the positive signal from the 
diffuse emission, which increasingly contributes to $S_{out}$ as $\sigma_{bck}$ rises 
(because less diffuse emission is removed in the background subtraction stage).
By $\sigma_{bck}/\sigma_{beam}=8$, the 84th percentile of $S_{out}/S_{in}$ is 
$\sim1.2$, compared to the 50th percentile of $\sim1.05$; the fluctuation towards
larger values of $S_{out}/S_{in}$ is therefore $\sigma_{+}=0.15$ or $15\%$ higher
than the input value.
It is likely that this effect is also present in Sim2, where a diffuse component is
present and sources are randomly distributed, but since far fewer sources are located
close to the galactic plane, where the background is brighter, the effect is less obvious.

The choice of $\sigma_{bck}$ is therefore a compromise between loss of source 
flux due to excessive background subtraction (smaller $\sigma_{bck}$), and
increasing contamination from diffuse emission (larger $\sigma_{bck}$).
Adopting $\sigma_{bck}/\sigma_{beam}=2.5$ as in the QUaD survey data results 
in $<5\%$ systematic loss of source flux, while reducing the scatter due to 
the background to $\sim5\%$ --- both these effects are comparable to the absolute 
calibration uncertainty in the maps of $3.5\%$.

\end{appendix}

\newpage
\begin{landscape}
\scriptsize
\begin{center}
\begin{longtable}{lll|cccc|cccc|cc|cc}
\caption{QUaD Galactic Source Catalog\label{tab:srccat}}\\

\hline \hline \\[-2ex]
\multicolumn{1}{c}{} &
\multicolumn{1}{c}{} &
\multicolumn{1}{c}{} &
\multicolumn{4}{c}{100 GHz} &
\multicolumn{4}{c}{150 GHz} &
\multicolumn{1}{c}{} &
\multicolumn{1}{c}{} &
\multicolumn{1}{c}{} &
\multicolumn{1}{c}{} \\[0.5ex] 
\multicolumn{1}{c}{\textbf{Source}} &
\multicolumn{1}{c}{\textbf{RA}} &
\multicolumn{1}{c}{\textbf{dec}} &
\multicolumn{1}{c}{\textbf{$\sigma_{maj}$}} &
\multicolumn{1}{c}{\textbf{$\sigma_{min}$}} &
\multicolumn{1}{c}{\textbf{$I$}} &
\multicolumn{1}{c}{\textbf{$\delta I$}} &
\multicolumn{1}{c}{\textbf{$\sigma_{maj}$}} &
\multicolumn{1}{c}{\textbf{$\sigma_{min}$}} &
\multicolumn{1}{c}{\textbf{$I$}} &
\multicolumn{1}{c}{\textbf{$\delta I$}} &
\multicolumn{1}{c}{\textbf{$\alpha_{I}$}} &
\multicolumn{1}{c}{\textbf{$\delta\alpha_{I}$}} &
\multicolumn{1}{c}{\textbf{$Alt. Name$}} &
\multicolumn{1}{c}{\textbf{$WC$}} \\[0.5ex]
\multicolumn{1}{c}{} &
\multicolumn{1}{c}{(deg)} &
\multicolumn{1}{c}{(deg)} &
\multicolumn{1}{c}{(arcmin)} &
\multicolumn{1}{c}{(arcmin)} &
\multicolumn{1}{c}{(Jy)} &
\multicolumn{1}{c}{(Jy)} &
\multicolumn{1}{c}{(arcmin)} &
\multicolumn{1}{c}{(arcmin)} &
\multicolumn{1}{c}{(Jy)} &
\multicolumn{1}{c}{(Jy)} &
\multicolumn{1}{c}{} &
\multicolumn{1}{c}{} &
\multicolumn{1}{c}{} &
\multicolumn{1}{c}{} \\[0.5ex] \hline
\\[-1.8ex]
\endfirsthead

\multicolumn{15}{c}{{\tablename} \thetable{} -- Continued} \\ [0.5ex]
\hline \hline \\[-2ex]
\multicolumn{1}{c}{} &
\multicolumn{1}{c}{} &
\multicolumn{1}{c}{} &
\multicolumn{4}{c}{100 GHz} &
\multicolumn{4}{c}{150 GHz} &
\multicolumn{1}{c}{} &
\multicolumn{1}{c}{} &
\multicolumn{1}{c}{} &
\multicolumn{1}{c}{} \\[0.5ex] 
\multicolumn{1}{c}{\textbf{Source}} &
\multicolumn{1}{c}{\textbf{RA}} &
\multicolumn{1}{c}{\textbf{dec}} &
\multicolumn{1}{c}{\textbf{$\sigma_{maj}$}} &
\multicolumn{1}{c}{\textbf{$\sigma_{min}$}} &
\multicolumn{1}{c}{\textbf{$I$}} &
\multicolumn{1}{c}{\textbf{$\delta I$}} &
\multicolumn{1}{c}{\textbf{$\sigma_{maj}$}} &
\multicolumn{1}{c}{\textbf{$\sigma_{min}$}} &
\multicolumn{1}{c}{\textbf{$I$}} &
\multicolumn{1}{c}{\textbf{$\delta I$}} &
\multicolumn{1}{c}{\textbf{$\alpha_{I}$}} &
\multicolumn{1}{c}{\textbf{$\delta\alpha_{I}$}} &
\multicolumn{1}{c}{\textbf{$Alt. Name$}} &
\multicolumn{1}{c}{\textbf{$WC$}} \\[0.5ex]
\multicolumn{1}{c}{} &
\multicolumn{1}{c}{(deg)} &
\multicolumn{1}{c}{(deg)} &
\multicolumn{1}{c}{(arcmin)} &
\multicolumn{1}{c}{(arcmin)} &
\multicolumn{1}{c}{(Jy)} &
\multicolumn{1}{c}{(Jy)} &
\multicolumn{1}{c}{(arcmin)} &
\multicolumn{1}{c}{(arcmin)} &
\multicolumn{1}{c}{(Jy)} &
\multicolumn{1}{c}{(Jy)} &
\multicolumn{1}{c}{} &
\multicolumn{1}{c}{} &
\multicolumn{1}{c}{} &
\multicolumn{1}{c}{} \\[0.5ex] \hline
\\[-1.8ex]
\endhead

\hline
\endfoot
\\[-1.8ex] \hline \hline
\endlastfoot

253.61-0.18 & 124.08 & -35.56 & 21.7 & 3.5 & 9.38 & 7.86 & 3.8 & 1.8 & 1.75 & 11.19 & -3.66 & 14.04 & IRAS 08143-3521  & 0 \\ 
254.69+0.21 & 125.23 & -36.23 & 2.1 & 1.9 & 2.08 & 0.13 & 1.5 & 1.4 & 2.55 & 0.16 & 0.45 & 0.19 & IRAS 08189-3602  & 1 \\ 
261.65-2.08 & 128.06 & -43.23 & ... & ... & ... & ... & 1.8 & 1.4 & 0.96 & 0.18 & ... & ... & IRAS 08286-4251  & 0 \\ 
262.11-1.80 & 128.74 & -43.43 & ... & ... & ... & ... & 5.1 & 2.1 & 2.59 & 0.90 & ... & ... & none & 0 \\ 
259.95-0.03 & 128.90 & -40.64 & ... & ... & ... & ... & 2.0 & 1.7 & 1.22 & 0.32 & ... & ... & IRAS 08318-4020  & 0 \\ 
260.81+0.68 & 130.33 & -40.89 & ... & ... & ... & ... & 2.0 & 1.1 & 0.78 & 0.24 & ... & ... & none & 0 \\ 
263.63-0.53 & 131.41 & -43.86 & 2.6 & 2.3 & 2.19 & 0.15 & 2.1 & 1.7 & 4.08 & 0.16 & 1.35 & 0.17 & IRAS 08438-4340  & 1 \\ 
263.79-0.43 & 131.65 & -43.92 & ... & ... & ... & ... & 1.6 & 1.5 & 1.55 & 0.14 & ... & ... & IRAS 08432-4335  & 0 \\
\end{longtable}
\end{center}

\begin{center}
\begin{longtable}{lll|cccccc|cccccc|cccc|c}
\caption{QUaD Polarized Galactic Source Catalog\label{tab:polsrccat}}\\

\hline \hline \\[-2ex]
\multicolumn{1}{c}{} &
\multicolumn{1}{c}{} &
\multicolumn{1}{c}{} &
\multicolumn{6}{c}{100 GHz} &
\multicolumn{6}{c}{150 GHz} &
\multicolumn{1}{c}{} &
\multicolumn{1}{c}{} &
\multicolumn{1}{c}{} &
\multicolumn{1}{c}{} &
\multicolumn{1}{c}{} \\[0.5ex] 
\multicolumn{1}{c}{\textbf{Source}} &
\multicolumn{1}{c}{\textbf{RA}} &
\multicolumn{1}{c}{\textbf{dec}} &
\multicolumn{1}{c}{\textbf{$I$}} &
\multicolumn{1}{c}{\textbf{$\delta I$}} &
\multicolumn{1}{c}{\textbf{$P$}} &
\multicolumn{1}{c}{\textbf{$\delta P$}} &
\multicolumn{1}{c}{\textbf{$\phi$}} &
\multicolumn{1}{c}{\textbf{$\delta \phi$}} &
\multicolumn{1}{c}{\textbf{$I$}} &
\multicolumn{1}{c}{\textbf{$\delta I$}} &
\multicolumn{1}{c}{\textbf{$P$}} &
\multicolumn{1}{c}{\textbf{$\delta P$}} &
\multicolumn{1}{c}{\textbf{$\phi$}} &
\multicolumn{1}{c}{\textbf{$\delta \phi$}} &
\multicolumn{1}{c}{\textbf{$\alpha_{I}$}} &
\multicolumn{1}{c}{\textbf{$\delta\alpha_{I}$}} &
\multicolumn{1}{c}{\textbf{$\alpha_{P}$}} &
\multicolumn{1}{c}{\textbf{$\delta\alpha_{P}$}} &
\multicolumn{1}{c}{\textbf{$Alt. Name$}} \\[0.5ex]
\multicolumn{1}{c}{} &
\multicolumn{1}{c}{(deg)} &
\multicolumn{1}{c}{(deg)} &
\multicolumn{1}{c}{(Jy)} &
\multicolumn{1}{c}{(Jy)} &
\multicolumn{1}{c}{(Jy)} &
\multicolumn{1}{c}{(Jy)} &
\multicolumn{1}{c}{(deg)} &
\multicolumn{1}{c}{(deg)} &
\multicolumn{1}{c}{(Jy)} &
\multicolumn{1}{c}{(Jy)} &
\multicolumn{1}{c}{(Jy)} &
\multicolumn{1}{c}{(Jy)} &
\multicolumn{1}{c}{(deg)} &
\multicolumn{1}{c}{(deg)} &
\multicolumn{1}{c}{} &
\multicolumn{1}{c}{} &
\multicolumn{1}{c}{} &
\multicolumn{1}{c}{} &
\multicolumn{1}{c}{} \\[0.5ex] \hline
\\[-1.8ex]
\endfirsthead

\multicolumn{8}{c}{{\tablename} \thetable{} -- Continued} \\ [0.5ex]
\hline \hline \\[-2ex]
\multicolumn{1}{c}{} &
\multicolumn{1}{c}{} &
\multicolumn{1}{c}{} &
\multicolumn{6}{c}{100 GHz} &
\multicolumn{6}{c}{150 GHz} &
\multicolumn{1}{c}{} &
\multicolumn{1}{c}{} &
\multicolumn{1}{c}{} &
\multicolumn{1}{c}{} &
\multicolumn{1}{c}{} \\[0.5ex] 
\multicolumn{1}{c}{\textbf{Source}} &
\multicolumn{1}{c}{\textbf{RA}} &
\multicolumn{1}{c}{\textbf{dec}} &
\multicolumn{1}{c}{\textbf{$I$}} &
\multicolumn{1}{c}{\textbf{$\delta I$}} &
\multicolumn{1}{c}{\textbf{$P$}} &
\multicolumn{1}{c}{\textbf{$\delta P$}} &
\multicolumn{1}{c}{\textbf{$\phi$}} &
\multicolumn{1}{c}{\textbf{$\delta \phi$}} &
\multicolumn{1}{c}{\textbf{$I$}} &
\multicolumn{1}{c}{\textbf{$\delta I$}} &
\multicolumn{1}{c}{\textbf{$P$}} &
\multicolumn{1}{c}{\textbf{$\delta P$}} &
\multicolumn{1}{c}{\textbf{$\phi$}} &
\multicolumn{1}{c}{\textbf{$\delta \phi$}} &
\multicolumn{1}{c}{\textbf{$\alpha_{I}$}} &
\multicolumn{1}{c}{\textbf{$\delta\alpha_{I}$}} &
\multicolumn{1}{c}{\textbf{$\alpha_{P}$}} &
\multicolumn{1}{c}{\textbf{$\delta\alpha_{P}$}} &
\multicolumn{1}{c}{\textbf{$Alt. Name$}} \\[0.5ex]
\multicolumn{1}{c}{} &
\multicolumn{1}{c}{(deg)} &
\multicolumn{1}{c}{(deg)} &
\multicolumn{1}{c}{(Jy)} &
\multicolumn{1}{c}{(Jy)} &
\multicolumn{1}{c}{(Jy)} &
\multicolumn{1}{c}{(Jy)} &
\multicolumn{1}{c}{(deg)} &
\multicolumn{1}{c}{(deg)} &
\multicolumn{1}{c}{(Jy)} &
\multicolumn{1}{c}{(Jy)} &
\multicolumn{1}{c}{(Jy)} &
\multicolumn{1}{c}{(Jy)} &
\multicolumn{1}{c}{(deg)} &
\multicolumn{1}{c}{(deg)} &
\multicolumn{1}{c}{} &
\multicolumn{1}{c}{} &
\multicolumn{1}{c}{} &
\multicolumn{1}{c}{} &
\multicolumn{1}{c}{} \\[0.5ex] \hline
\\[-1.8ex]

\endhead
\multicolumn{8}{l}{{Continued on Next Page\ldots}} \\
\endfoot
\\[-1.8ex] \hline \hline
\endlastfoot

284.33-0.36 & 156.06 & -57.81 & 143.6 & 0.3 & 2.80 & 1.09 & 13.68 & 11.15 & 138.32 & 0.36 & ... & ... & ... & ... & -0.08 & 0.01 & ... & ... & IRAS 10227-5730  \\ 
359.93-0.06 & 266.42 & -29.03 & 175.2 & 0.4 & ... & ... & ... & ... & 171.03 & 0.46 & 1.08 & ... & 75.62 & ... & -0.05 & 0.01 & ... & ... & none \\ 
0.18-0.06 & 266.57 & -28.81 & ... & ... & 7.91 & 0.33 & -38.61 & 1.19 & ... & ... & 4.90 & 0.32 & -42.20 & 1.86 & ... & ... & -1.04 & 0.17 & IRAS 17431-2846  \\ 
0.63-0.05 & 266.82 & -28.42 & 112.7 & 0.3 & 3.21 & 7.35 & 88.40 & 65.64 & 139.83 & 0.27 & 1.33 & 0.24 & 68.57 & 5.19 & 0.47 & 0.01 & -1.92 & 5.00 & IRAS 17440-2823  \\
\end{longtable}
\end{center}

\normalsize
\newpage
\end{landscape}

\newpage

\newpage

\end{document}